\documentclass[aip,jcp,footinbib,citeautoscript,preprint]{revtex4-1}
\usepackage{graphicx} 
\usepackage{amsmath}
\usepackage{amssymb} 
\usepackage[dvipsnames]{xcolor}
\usepackage{hyperref}
\usepackage{soul}

\usepackage{braket} 

\newcommand{\eu}{\mathrm{e}^}
\newcommand{\iu}{\ensuremath{\mathrm{i}}}
\newcommand{\rmd}{\mathrm{d}}
\newcommand{\thalf}{{\ensuremath{\tfrac{1}{2}}}} 
\newcommand{\half}{{\ensuremath{\frac{1}{2}}}}

\DeclareMathOperator{\Real}{Re}
\DeclareMathOperator{\Imag}{Im}
\renewcommand{\Re}{\Real}
\renewcommand{\Im}{\Imag}

\newcommand{\eqn}[1]{Eq.\,\eqref{#1}}
 \newcommand{\fig}[1]{Fig.\,\ref{fig:#1}}
\newcommand{\Ref}[1]{Ref.~\onlinecite{#1}}
\newcommand{\sref}[1]{Sec.~\ref{sec:#1}\@}

\usepackage{dcolumn}
\newcolumntype{d}[1]{D{.}{.}{#1}} 

\DeclareMathOperator{\csch}{csch}

\usepackage{pdfpages}

\makeatletter
\AtBeginDocument{\let\LS@rot\@undefined}
\makeatother

\begin{document}

\title{Nonadiabatic quantum transition-state theory in the golden-rule limit. II.
Overcoming the pitfalls of the saddle-point and semiclassical approximations
}

\author{Wei Fang}
\author{Manish J. Thapa}
\author{Jeremy O. Richardson}
\email{jeremy.richardson@phys.chem.ethz.ch}
\affiliation{Laboratory of Physical Chemistry, ETH Zurich, 8093 Zurich, Switzerland}

\date{\today}

\begin{abstract}
\noindent
We describe a path-integral molecular dynamics implementation of our recently developed golden-rule quantum transition-state theory (GR-QTST).
The method is applied to compute the reaction rate in various models of electron transfer and benchmarked against exact results.
We demonstrate that for systems exhibiting two or more transition states,
rates computed using Wolynes theory [P. G. Wolynes, J.\ Chem.\ Phys.\ 87, 6559 (1987)]
can be overestimated by orders of magnitude,
whereas the GR-QTST predictions are numerically accurate.
This is the case both at low temperature, where nuclear tunneling makes a considerable contribution,
and also in the classical limit, where only GR-QTST rigorously tends to the correct result.
Analysis shows that the saddle-point approximation employed by Wolynes theory is not valid in this case,
which results in predictions of unphysical reaction pathways,
whilst the energy constraint employed by GR-QTST resolves this problem.
The GR-QTST method is also seen to give accurate results
for a strongly anharmonic system
by sampling configurations around the instanton pathway without making the semiclassical approximation.
These promising results indicate that the GR-QTST method could be an efficient and accurate approach
for simulating electron-transfer reactions in complex molecular systems.
\end{abstract}

\maketitle

%%%%%%%%%%%%%%%%%%%%%%%%%%%%%%%%%%%%%%%%%%

\section{Introduction}
Electron transfer is one of the fundamental mechanisms of chemical reactions
with relevance to biological processes, electrochemical tunneling microscopy and spectroscopy, and industrial applications \cite{Blumberger2015chemrev,Friis1379,doi:10.1021/nl103334q,doi:10.1021/nn103236e}.
Such reactions can be described by a system of two weakly-coupled electronic states
interacting with a large number of nuclear degrees of freedom.\cite{ChandlerET}
Electron-transfer reactions typically take place in the golden-rule limit 
and one cannot therefore employ the Born--Oppenheimer approximation.
Instead, the rate is formally given by
Fermi's golden-rule 
\cite{Fermi}.
In practice, however, further approximations are also required for a system in which the exact nuclear wave functions
cannot be calculated.
Marcus theory \cite{Marcus1964review,Marcus1993review}, which
effectively treats the free-energy curves harmonically and neglects nuclear quantum effects,
has been widely applied to model these reactions.
Various methods have been developed to improve simultaneously upon both of these approximations of Marcus theory
\cite{%
%Siders1981quantum,
Wolynes1987nonadiabatic,
Wolynes1987dissipation,*Cline1987nonadiabatic,*Onuchic1988rate,
Lawrence2018Wolynes,
Zhu1994ZN,
%Rips1995ET,
Cao1995nonadiabatic,*Cao1997nonadiabatic,*Schwieters1998diabatic,*Schwieters1999diabatic,
HammesSchiffer2010PCET,*HammesSchiffer2015PCET,
%Menzeleev2011ET, Shushkov2013instanton, % these use classical solvet
Kretchmer2013ET, % extends Menzeleev2011ET to have RPMD proton too
Menzeleev2014kinetic,Kretchmer2016KCRPMD,
%Kretchmer2018KCRPMD, % here they don't quantize nuclei
%nonoscillatory,
GoldenGreens,GoldenRPI,AsymSysBath,
%Skinner1997golden,
%Wang1999mapping,
Duke2016Faraday,*Pierre2017MVRPMD,
%Huo2013PLDM,
%Tully1990hopping,
%Shushkov2012RPSH,
%Landry2012hopping,*Jain2015hopping2,
Tao2019RPSH,
doi:10.1063/1.4803835,PhysRevLett.105.123002,doi:10.1021/acs.jpclett.7b01249,
doi:10.1063/1.4981021,doi:10.1021/acs.jpclett.7b01343%
%Spencer2016Faraday,%Tiwari2016Faraday,
%others?
%GRQTST % should we include this here or only later?
}
and
several studies have revealed the importance of quantum effects such as nuclear tunneling, even at room temperature
\cite{
Bader1990golden,
Siders1981inverted,AsymSysBath,
HammesSchiffer2008PCET,Hu2014enzyme,Pollak2012tunnel%
%Menzeleev2014kinetic,Kretchmer2013ET
}.

We will concentrate on methods based on the path-integral description of quantum mechanics \cite{Feynman}
which are able to describe nuclear tunneling effects.
One such approach is semiclassical instanton rate theory in the golden-rule limit, \cite{GoldenGreens}
which is an extension of the original instanton theory appropriate in the Born--Oppenheimer limit
\cite{Miller1975semiclassical,Uses_of_Instantons,Althorpe2011ImF}.
Both theories can be derived in a similar way 
using a number of steepest-descent approximations to the path-integral formulation of the flux correlation function
to obtain a formula for the rate
defined in terms of imaginary-time periodic classical trajectories and harmonic fluctuations around them.
\cite{GoldenGreens,AdiabaticGreens,InstReview}
Each of these periodic trajectories is called an ``instanton'', and together they define a set of optimal tunneling pathways,
each of which is identified as a mechanism of the reaction.
Like the original version \cite{Andersson2009Hmethane,RPInst,Perspective},
the golden-rule instanton theory can also be efficiently evaluated in the ring-polymer formulation \cite{GoldenRPI},
and has been benchmarked on system-bath models and shown to give accurate predictions for electron-transfer rates \cite{AsymSysBath}.
However, although the theory is applicable to anharmonic systems,
as the semiclassical limit only treats fluctuations to second order,
it is equivalent to performing a local harmonic approximation,
which cannot be used successfully in atomistic solvent environments.
Wolynes theory, \cite{Wolynes1987nonadiabatic} on the other hand,
does not make a semiclassical approximation
and instead evaluates the path integral numerically by sampling discretized paths, known as ring polymers. \cite{Chandler+Wolynes1981}
It can therefore be applied to atomistic environments using techniques developed for classical molecular dynamics.
\cite{Parrinello1984Fcenter,Markland2018review}
However, only imaginary-time (statistical) properties are rigorously available from this approach
and so an expression for the rate is derived
by integrating out the real-time behaviour
using a saddle-point approximation.
The method has been employed to compute electron-transfer rates in a number of applications. \cite{Zheng1989ET,*Zheng1991ET,Bader1990golden}

In \Ref{GRQTST} (hence referred to as Paper I),
we proposed a new golden-rule quantum transition-state theory (GR-QTST) for predicting rates of electron-transfer reactions.
This method was designed to go beyond instanton theory
by sampling paths around the instanton
in a similar way to ring-polymer transition-state theory \cite{RPInst,Hele2013QTST,Mills1997QTST}
and the projected quantum instanton method \cite{QInst}.
It also extends the ideas of classical golden-rule transition-state theory \cite{ChandlerET,nonoscillatory},
in which energy conservation from the reactant to product states is enforced by a constraint.
The method thus 
differs from Wolynes theory in that a constraint is applied to every path in the ensemble.
In this respect, there is some similarity to the kinetically-constrained ring-polymer molecular dynamics (KC-RPMD) method \cite{Menzeleev2014kinetic,Kretchmer2016KCRPMD,Kretchmer2018KCRPMD},
although there are important differences in the definition and implementation.

In the previous work, \cite{GRQTST} we tested the GR-QTST method on a spin-boson model and a one-dimensional anharmonic system
and found that the results were consistently in good agreement with the exact rate
and were also more accurate than published KC-RPMD results \cite{Menzeleev2014kinetic,Kretchmer2016KCRPMD}.
In addition, we tested other approaches including instanton theory and Wolynes theory,
which were also seen to perform well.
In fact, Wolynes and instanton theories typically gave slightly more accurate results than GR-QTST in these cases
(except for a particular discrepancy of Wolynes theory, which does not tend to the correct classical limit for high temperatures in an anharmonic system).
However, we will show in \sref{results}
that these two methods break down
when applied to more complex systems than the simple models employed in Paper I,
leaving GR-QTST as the only consistently reliable method.

In this work, 
we argue that the saddle-point approximation of Wolynes theory is not an asymptotic approximation in $\hbar$,
and for this reason does not rigorously tend to the correct classical limit.
We present a simple example where it completely breaks down by introducing a system with two transition states in \sref{theory}.
Here, transition states (TSs) are defined loosely
from a classical viewpoint
as a point of minimum energy on the crossing seam between two diabatic surfaces. \cite{nonoscillatory}
We note that Wolynes himself already hinted at the possibility of a break-down in such systems.\cite{Wolynes1987nonadiabatic}
By ``break down'' we mean that the method overestimates or underestimates the rate by at least an order of magnitude.
A rate method liable to make this sort of error will also typically lose the ability to correctly identify the reaction mechanism,
and so we will also evaluate the utility of the various methods with respect to this important criterion.

In order to perform the GR-QTST calculations presented in \sref{results},
we introduce a path-integral molecular dynamics (PIMD) implementation in \sref{methods}.
For the strongly anharmonic systems employed in this work,
this approach is more efficient than the importance sampling method employed in Paper I
and will in principle allow the method to be applied to atomistic simulations.

It is clear that the semiclassical approximation of instanton theory breaks down when the fluctuations around the
instanton pathway become strongly anharmonic,
such as in a system with explicit solvents.
It was exactly for this reason that we developed the GR-QTST method,
and we demonstrate in \sref{anharmonic} that the new approach remains accurate even when instanton theory fails.
Finally, we conclude in \sref{conclusions} that GR-QTST successfully overcomes the pitfalls
of both the saddle-point and semiclassical approximations.

\section{Theory and analysis}
\label{sec:theory}
In this section, we discuss semiclassical instanton theory and Wolynes theory,
two methods derived in different ways from saddle-point approximations.
The limitations of instanton theory due to its harmonic treatment of path fluctuations are well understood, but Wolynes theory has not been subjected to the same scrutiny.
In particular, we will explain how the saddle-point approximation employed in the derivation of Wolynes theory
can lead to errors greater than an order of magnitude in the rate predictions,
even if the system is in the classical limit.
We then show how the energy constraint introduced in the GR-QTST method corrects for the problems of both methods
such that accurate results are obtained.

Consider an electron-transfer process from the reactant diabatic state $|0\rangle$ to the product diabatic state $|1\rangle$. \cite{ChandlerET}
The Hamiltonian is
\begin{equation}
\hat{H}=\hat{H}_0|0\rangle\langle0| + \hat{H}_1|1\rangle\langle1| + \Delta(|0\rangle\langle1|+|1\rangle\langle0|),
\end{equation}
where $\Delta$ is the electronic coupling,
and $\hat{H}_n = \hat{p}^2/2m + V_n(\hat{x})$ is a one-dimensional nuclear Hamiltonian for the electronic state $\ket{n}$
with potential-energy surface $V_n(x)$.
The nuclear coordinate is $x$ which has associated mass $m$ and conjugate momentum $p$.
Because our methods are based on a path-integral formalism, they are easily extended to treat
a system with $D$ nuclear degrees of freedom with
nuclear coordinates $\mathsf{x}=(x_1,\dots,x_D)$ and conjugate momenta $\mathsf{p}=(p_1,\dots,p_D)$, each with an associated mass $m_j$.
For these multidimensional systems, the nuclear Hamiltonian for each diabatic state is
$\hat{H}_n = \sum_{j=1}^D \hat{p}_j^2/2m_j + V_n(\hat{\mathsf{x}})$. 
Here we will assume that $\Delta$ is a constant, which is known as the Condon approximation, 
but all the methods we present are easily generalized to allow for a position-dependent coupling.

The quantum rate constant from state $|0\rangle$ to $|1\rangle$ can be defined
as the integral of a flux correlation function over time.
\cite{Yamamoto1960rate,Miller1974QTST,Miller1983rate,Wolynes1987nonadiabatic,ChandlerET}
We shall assume that the reaction proceeds in the nonadiabatic golden-rule limit,
i.e.\ $\Delta$ is small enough such that only the leading term need be considered for the rate.
\footnote{The simplest way to go beyond this limit would be to use the approach described in Ref.~\onlinecite{Lawrence2019ET}}\nocite{Lawrence2019ET}
The correlation function then takes on a particularly simple form and
leads to an expression for 
Fermi's golden-rule rate constant 
given by \cite{Weiss}
\begin{align}
\label{k}
	k Z_0=\frac{\Delta^2}{\hbar^2} \int_{-\infty}^{\infty} \eu{-\phi_\text{u}(\tau-\iu t)/\hbar} \, \rmd t,
\end{align}
where $Z_0=\textrm{Tr}_\textrm{n}[\eu{-\beta\hat{H}_0}]$ is the reactant partition function
and
the \emph{unconstrained} effective action, $\phi_\text{u}(\tau-\iu t)$, is defined by the correlation function
\begin{align}
\label{exact}
 \eu{-\phi_\text{u}(\tau-\iu t)/\hbar}=\textrm{Tr}_\textrm{n}[\eu{ -\hat{H}_0 (\tau - \iu t)/\hbar} \, \eu{- \hat{H}_1(\beta \hbar-\tau + \iu t)/\hbar}].
\end{align}
The symbol $\textrm{Tr}_\textrm{n}$ is a trace over nuclear degrees of freedom only.
Typically this trace cannot be evaluated exactly.
However, in this section, we will illustrate the theoretical concepts using a simple model system where the analytical result can be found
(Appendix~\ref{sec:1danalytic}).

It should be noted that this exact expression for the rate is independent of the choice of $\tau$. 
This is because
$\phi_\text{u}(z)$ is an 
analytic function 
of the complex variable $z$,
where $\tau=\Re z$ and $t=-\Im z$.
Therefore the contour integral can be deformed to effectively change the value of $\tau$ without affecting the result of the integral.
Finally, note that $\phi_\text{u}(z)$ is a real number when $t=0$, but is complex in general.
In general, the major difficulty in evaluating the integral in \eqn{k} numerically by quadrature is
because the integrand itself (and its first few derivatives) can only be obtained at short times,
as imaginary-time path-integral sampling techniques only allow an efficient numerical computation for $t=0$. \cite{Chandler+Wolynes1981}
We will discuss two methods in which
this integral is carried out by the method of steepest descent, which is also known as a saddle-point approximation. 
\cite{BenderBook}

The semiclassical instanton rate \cite{GoldenGreens}
is obtained by replacing the two propagators in \eqn{exact} by their semiclassical limits
(a sum over imaginary-time classical paths) \cite{GutzwillerBook}
and evaluating the trace over nuclear degrees of freedom by steepest-descent integration.
We call this the semiclassical approximation because it is exact in the limit $\hbar\rightarrow0$
(keeping the total imaginary time, $\beta\hbar$, fixed). 
Finally a further steepest-descent approximation is taken for the time integral. 
Alternatively, one can perform the steepest-descent integrals over path variables and time simultaneously,
leading to the same result. \cite{InstReview}
The stationary points in this extended space correspond to instanton configurations,
each with $t=0$ but a different value of $\tau$.
Taking the high-temperature, large-mass limit gives a formula \cite{GoldenGreens}
which is recognized as a local harmonic approximation to classical golden-rule transition-state theory. \cite{nonoscillatory,ChandlerET}

Wolynes theory \cite{Wolynes1987nonadiabatic}
does not employ the semiclassical approximation (steepest-descent integration over the path variables) to compute
the effective action, but does
employ a saddle-point approximation for the integral over time in \eqn{k}.
This approximation is equivalent to expanding $\phi_\text{u}(z)$ as a Taylor series to second order about its saddle point, 
thus effectively assuming that the correlation function has a Gaussian form centred on $t=0$, whose integral is known.
This is commonly called a second-order cumulant expansion.
The resulting expression for the rate is formulated as
\begin{align}
\label{wolk}
 k_\textrm{Wolynes} Z_0 = \frac{\Delta^2}{\hbar^2}\sqrt{2 \pi \hbar} \bigg[-\frac{\rmd^2 \phi_\text{u}}{\rmd \tau^2} \bigg]^{-\frac{1}{2}}_{\tau=\tau^*} \eu{-\phi_\text{u}(\tau^*)/\hbar},
\end{align} 
where $\tau^*$ is a real number and is defined as the stationary point which maximizes the unconstrained effective action, $\phi_\text{u}(\tau)$. 
\footnote{Typically $\tau^*$ is in the range $[0,\beta\hbar]$, but $\phi_\text{u}(\tau)$ can be analytically continued beyond this range to treat the inverted regime as described in \Ref{Lawrence2018Wolynes}}
This approximation has been called a 
saddle-point approximation, \cite{Wolynes1987nonadiabatic} but note that is not 
an asymptotic approximation \cite{BenderBook} in $\hbar$ 
as $\phi_\text{u}(\tau-\iu t)$ itself depends on $\hbar$.
This is in contrast to the steepest-descent integration used to derive semiclassical instanton theory,
which is a rigorous asymptotic approximation \cite{AdiabaticGreens,Faraday,InstReview}
and thus becomes exact in the limit $\hbar\rightarrow0$.

As pointed out in \Ref{nonoscillatory} and demonstrated numerically in Paper I,
the Wolynes rate does not tend correctly to the classical limit at very high temperatures
(unless the system is harmonic).
Nonetheless, the errors in previous examples were rather small.
In this paper, we demonstrate that orders-of-magnitude error can result from its saddle-point approximation,
in particular for systems with more than one transition state.
We shall investigate a system which exhibits two diabatic crossings,
denoted as A and B\@.
If we assume that the two crossing points are well separated in coordinate space, or couple different electronic states,
the effective action is
\begin{align}
\label{distribution}
 \eu{-\phi_\text{u}(\tau+\iu t)/\hbar}= \eu{-\phi_\text{u}^{(\text{A})}(\tau+\iu t)/\hbar}+ \eu{-\phi_\text{u}^{(\text{B})}(\tau+\iu t)/\hbar}.
\end{align}
This expression is analogous to forming a joint probability distribution by 
algebraically adding two independent probability distributions.
The exact rate expression can therefore be broken into two parts, 
$k=k^\text{(A)}+k^\text{(B)}$,
where $k^{(s)}$ is the rate through one of the crossings $s\in\{\text{A},\text{B}\}$.
Because the integration of a sum is simply the sum of integrals,
it can be shown that the classical and semiclassical instanton rates 
are rigorously separable in this same way.
It is also clear that this same argument can be generalized to systems with more than two crossings.

In contrast, Wolynes theory is not separable
and the rate obtained from one single calculation is different from the sum of two separate calculations.
In fact, if we apply Wolynes theory directly to the total effective action, $\phi_\text{u}(\tau)$,
the saddle-point approximation may be of very poor quality.
This is in spite of the fact that the sum of two separate Wolynes theory calculations
on the individual crossings may give a much better approximation to the rate.
We call this modified method ``separated Wolynes theory''.

\begin{figure}[!ht]
    \centering
    \includegraphics[width=8.5cm]{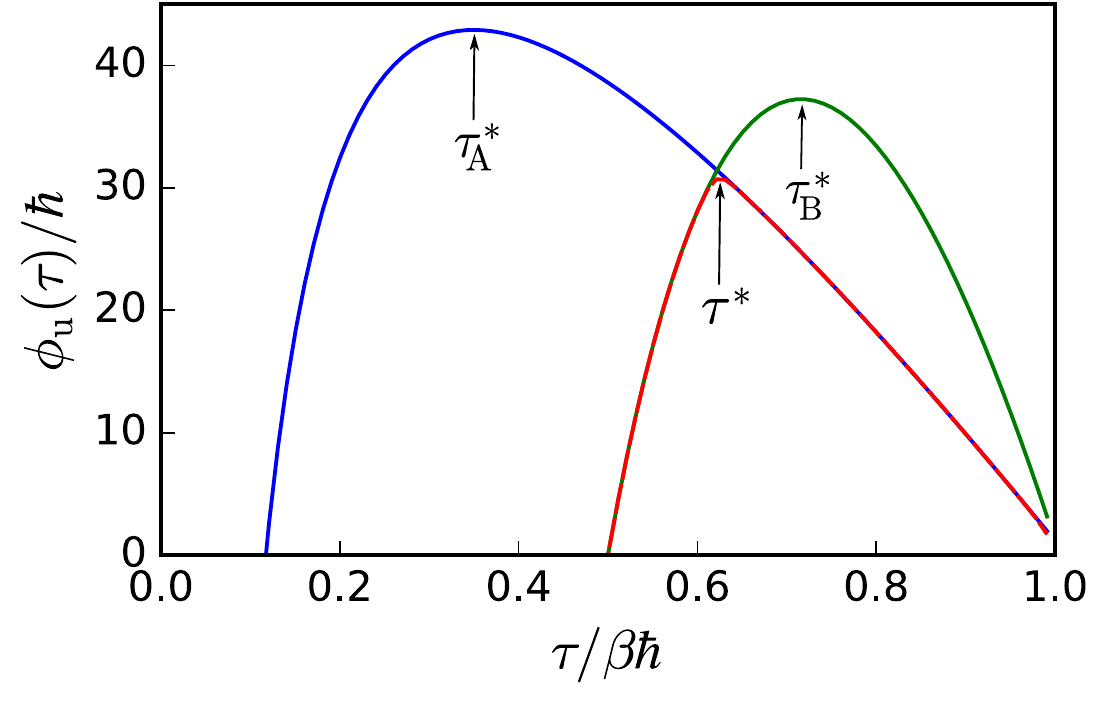}
    \caption{The dependence of the unconstrained effective action on $\tau$, with $\beta\hbar$ fixed.
	 The two $\phi_\text{u}^{(s)}(\tau)$ curves are evaluated separately using Eq.~\eqref{phi1d} for the transition states corresponding to A (blue) and B (green), 
	 whereas $\phi_\text{u}(\tau)$ (dashed red) is the combined result, defined by \eqn{distribution}.
	 }
    \label{phiplot}
\end{figure}
We demonstrate these ideas by considering the specific example of a system with a harmonic reactant potential
weakly coupled to two product states described by linear potentials which make crossings at the same height but with different slopes.
The parameters are defined in Appendix~\ref{sec:1danalytic} along with analytic expressions for the unconstrained effective action.
Fig.~\ref{phiplot} shows plots of 
$\phi_\text{u}^{(\text{A})}(\tau)$ and $\phi_\text{u}^{(\text{B})}(\tau)$
which are compared to $\phi_\text{u}(\tau)$,
defined by \eqn{distribution} with $t=0$.
It is seen that each of these three action functions has a maximum at a different value of $\tau$.

\begin{figure}[!ht]
    \centering
    \includegraphics[width=7.5cm]{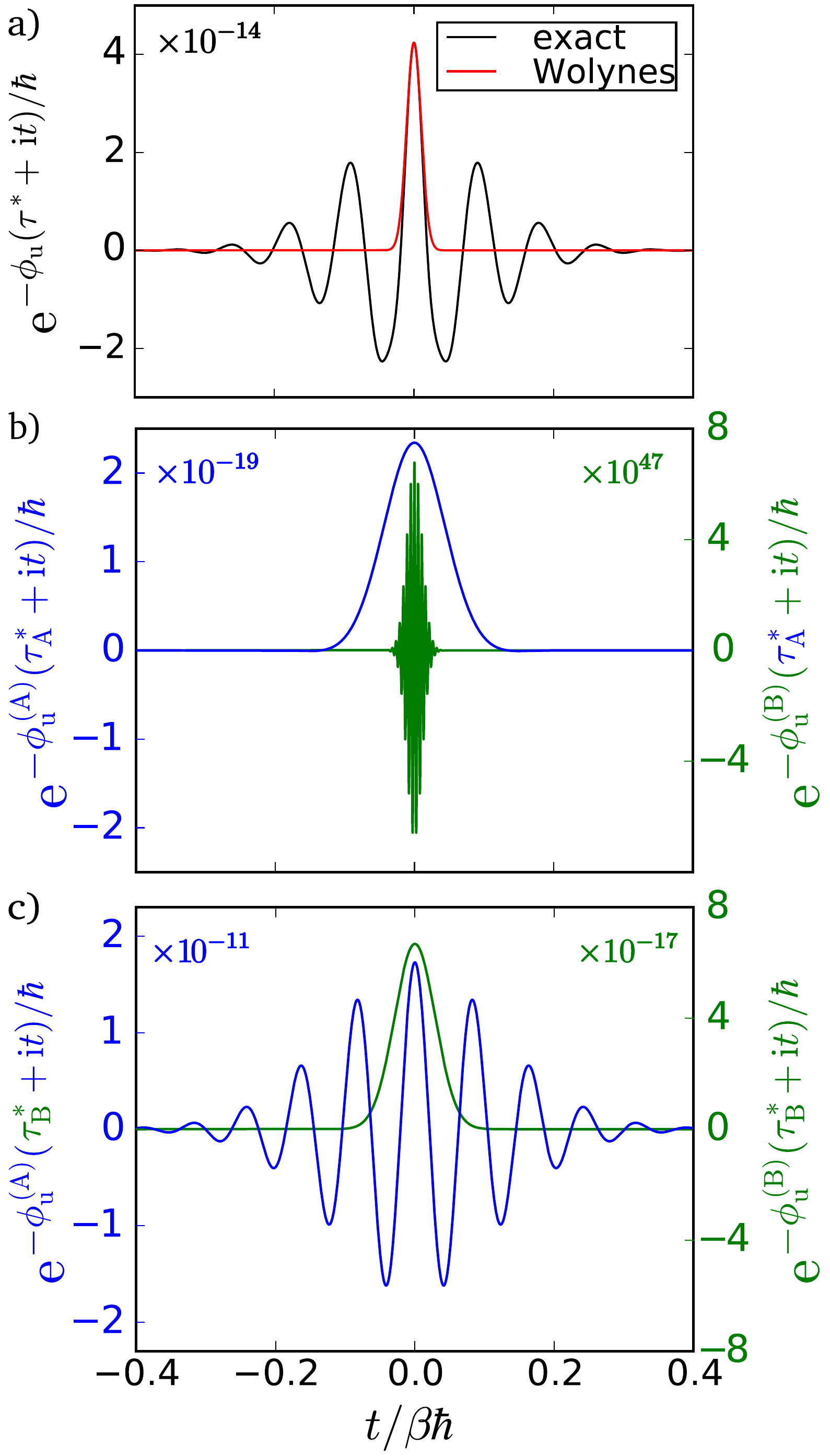}
    \caption{
	 The correlation functions, \eqn{exact}, computed with various values of imaginary time (only the real parts are shown).
	 Panel (a) shows the exact result compared with the saddle-point approximation of Wolynes theory using $\tau^*/\beta \hbar \approx 0.63$.
	 Plots in (b) and (c) show the exact results using $\tau_{s}^*$ for each of the transition states, $s \in \{\text{A},\text{B}\}$.
	 For transition state A (blue), $\tau^*_\text{A}/\beta \hbar \approx 0.35$, while for transition state B (green), $\tau^*_\text{B}/\beta \hbar \approx 0.72$.
	 Note the different scales used for the two plots using the colour-coded left or right y-axis.
    }
    \label{cff_final}
\end{figure}

Therefore, following the original Wolynes theory, \cite{Wolynes1987nonadiabatic}
one would obtain a single value of $\tau^*$ at the maximum of $\phi_\text{u}(\tau)$
and employ a saddle-point approximation around this value to yield the rate.
However, this does not lead to a good approximation
because, as is shown in Fig.~\ref{cff_final}(a),
the correlation function for this system defined with $\tau=\tau^*$ 
can be strongly oscillatory.
In principle, the exact rate can be obtained as the integral this correlation function,
but the cumulant expansion employed by Wolynes theory attempts to approximate it by a Gaussian.
The naive saddle-point approximation is therefore clearly not valid in the usual sense of a steepest-descent approximation,
which is supposed to remove the imaginary component of the exponent, \cite{BenderBook}
but here does so only for a very small region around the origin.
Wolynes theory, which can only approximate the positive part of the first peak,
can thus overestimate the integral by many orders of magnitude.

The correlation functions for the two transition states can also be investigated separately.
The exact total rate is given by the sum of the integrals over one blue and one green curve chosen from either Fig.~\ref{cff_final}(b) or (c).
One can see that 
either of the $\eu{-\phi_\text{u}^{(s)}(\tau+\iu t)/\hbar}$ correlation functions
can be made approximately Gaussian,
but that different choices of $\tau^*_s$ are required for this,
which cannot be simultaneously satisfied.
The choice $\tau=\tau^*$ made by Wolynes theory
removes neither the oscillations of $\eu{-\phi_\text{u}^\text{(A)}(\tau+\iu t)}$ nor of $\eu{-\phi_\text{u}^\text{(B)}(\tau+\iu t)}$.
The semiclassical approximation of instanton theory
follows a different approach
and integrates over the blue part of (b) and the green of (c), both of which can be well approximated by Gaussians.
In principle Wolynes theory could be systematically improved by going to higher-order cumulant expansions,
but this would not be practical because a huge number of terms would be required to describe the oscillatory behaviour.
It is therefore clear that in this case, the semiclassical approximation is superior to the cumulant expansion.
However, note that this may not be the case for strongly anharmonic systems such as those we consider in \sref{results}.

A physical interpretation for
the derivative of the effective action is a measure of the difference in energy between reactant and product states.
Thus choosing $\tau=\tau_s^*$ at the stationary point of $\phi_\text{u}^{(s)}(\tau)$ ensures that the energies match around this transition state, as is required by Fermi's golden rule. \cite{nonoscillatory}
However, 
the value of $\tau=\tau^*$ which maximizes $\phi_\text{u}(\tau)$ 
matches the average reactant energy to the average product energy globally,
but they are not necessarily matched locally around each transition state.
Wolynes theory will therefore sample unphysical paths that do not resemble instanton-like configurations,
leading to a loss of mechanistic insight as well as
orders-of-magnitude error in the predicted rate.
We can also make some statements about the effect of this problem in more general systems.
Firstly, assuming that both $\phi_\text{u}^{(s)}$ curves exhibit only one maximum each and that $\tau_\text{A}^*<\tau_\text{B}^*$,
then $\tau^*$, which is the maximum of $\phi_\text{u}(\tau)$, will always be in between the two maxima, i.e.\ $\tau_\text{A}^* < \tau^* < \tau_\text{B}^*$.
\footnote{This can be proved by setting the derivative of \eqn{distribution} to 0, which requires the derivatives of $\phi_\text{u}^{(\text{A})}(\tau)$ and $\phi_\text{u}^{(\text{B})}(\tau)$ to have opposite signs.}
Also, $\phi_\text{u}(\tau^*)$ will always be smaller than either of $\phi_\text{u}^{(s)}(\tau_s^*)$.
Therefore, although Wolynes theory is not variational in general, the problem we identify here will typically cause the rate to be overpredicted. 
However, it will only have a significant impact in cases for which
$\phi_\text{u}^{(\text{A})}(\tau)=\phi_\text{u}^{(\text{B})}(\tau)$ at a value of $\tau$ in between $\tau_\text{A}^*$ and $\tau_\text{B}^*$.

Of course, for this simple model system, one can fix the problem by computing the rates using the separated Wolynes method,
in which the saddle-point approximation
is evaluated independently for the two transition states at $\tau_\text{A}^*$ and $\tau_\text{B}^*$.
The total rate is given by the sum of these two rates.
Unfortunately it is not always so easy to separate the different contributions to Wolynes theory,
in particular for complex molecular environments.
It is for this reason we have proposed the GR-QTST method, \cite{GRQTST}
which we have developed to avoid this problem.

The results from various rate calculations are presented in Table \ref{table:1dtable},
where they can be compared with the benchmark exact result from Fermi's golden-rule.
The difference between classical and exact rates for this system indicates that we are in the quantum regime where tunneling is responsible for enhancing the rates by over three orders of magnitude.
It is also worth noting that due to its neglect of tunneling, classical mechanics incorrectly predicts a faster rate through transition state A\@.
Semiclassical instanton rates, however, are very accurate,
which is expected for this system which includes no anharmonicities such that the semiclassical approximation is exact for integrals over nuclear degrees of freedom.
Therefore, 
there is only a very minor difference between semiclassical instanton theory and the separated Wolynes theory here.
In both cases the only
approximation used is to employ the saddle-point method to the time integral over each individual Gaussian-like correlation function.
Instanton theory is calculated separately at $\tilde{\tau}_s$, which is the maximum of the classical action $S^{(s)}(\tau)$ [\eqn{Su}], and separated Wolynes theory at $\tau_s^*$, which is the maximum of $\phi_\text{u}^{(s)}(\tau)$ [\eqn{phi1d}].
For this system, we find that $\tilde{\tau}_s$ and $\tau_s^*$ differ by less than 1\% at either of the transition states,
thus giving almost identical predictions from the two theories.

\begin{table*}
\caption{Results of classical [Eq.~\eqref{LZ}], exact [Eq.~\eqref{k}], semiclassical (SC) instanton [Eq.~\eqref{instphi}], GR-QTST [Eq.~\eqref{grqtst}] and Wolynes calculations [Eq.~\eqref{wolk}]
for the one-dimensional system defined by \eqn{AnharmonicSys}.
$N=200$ ring-polymer beads were used to perform the GR-QTST calculations with
5000 Monte-Carlo steps.
The statistical errors of one standard deviation are given.
Instanton and Wolynes rate calculations have been performed analytically using $\phi_\text{u}^{(s)}(\tau)$ obtained in Eq.~\eqref{phi1d}.
Comparison between an instanton calculation with a finite number of beads and the analytic result shows that $N=200$ is sufficient to describe the tunneling accurately.
To make the results dimensionless, all the rates have been divided by $k_{\text{cl}}$, which is the total classical rate for the reaction through both transition states.
}
\begin{ruledtabular}\begin{tabular}{ llll } 
 method  & \multicolumn{1}{c}{$k^{(\text{A})}/k_{\text{cl}}$} & \multicolumn{1}{c}{$k^{(\text{B})}/k_{\text{cl}}$} & \multicolumn{1}{c}{$k/k_{\text{cl}}$} \\
 \hline
classical  & \phantom{0}0.6667 & \phantom{000}0.3333 & \phantom{000}1\\ 
exact  & 10.64 & 2068 & 2079\\
Wolynes ($N\rightarrow\infty$)    &-&-&\!\!\!\!\!\!$471300$\\
Wolynes separated ($N\rightarrow\infty$)   & 10.70 & 2068 & $2079$\\
SC ($N\rightarrow\infty$) & 10.62 & 2067 & 2078\\ 
SC ($N=200$) & 10.62 & 2064 & 2075\\ 
GR-QTST ($N=200$)    & - & - &$2116 \pm4$\\ 
GR-QTST separated ($N=200$)  & $10.82\pm 0.02$ & $2119 \pm 4$ & $2130 \pm 4$\\ 
\end{tabular}\end{ruledtabular}
\label{table:1dtable}
\end{table*}

\begin{figure}
    \centering
    \includegraphics[width=8cm]{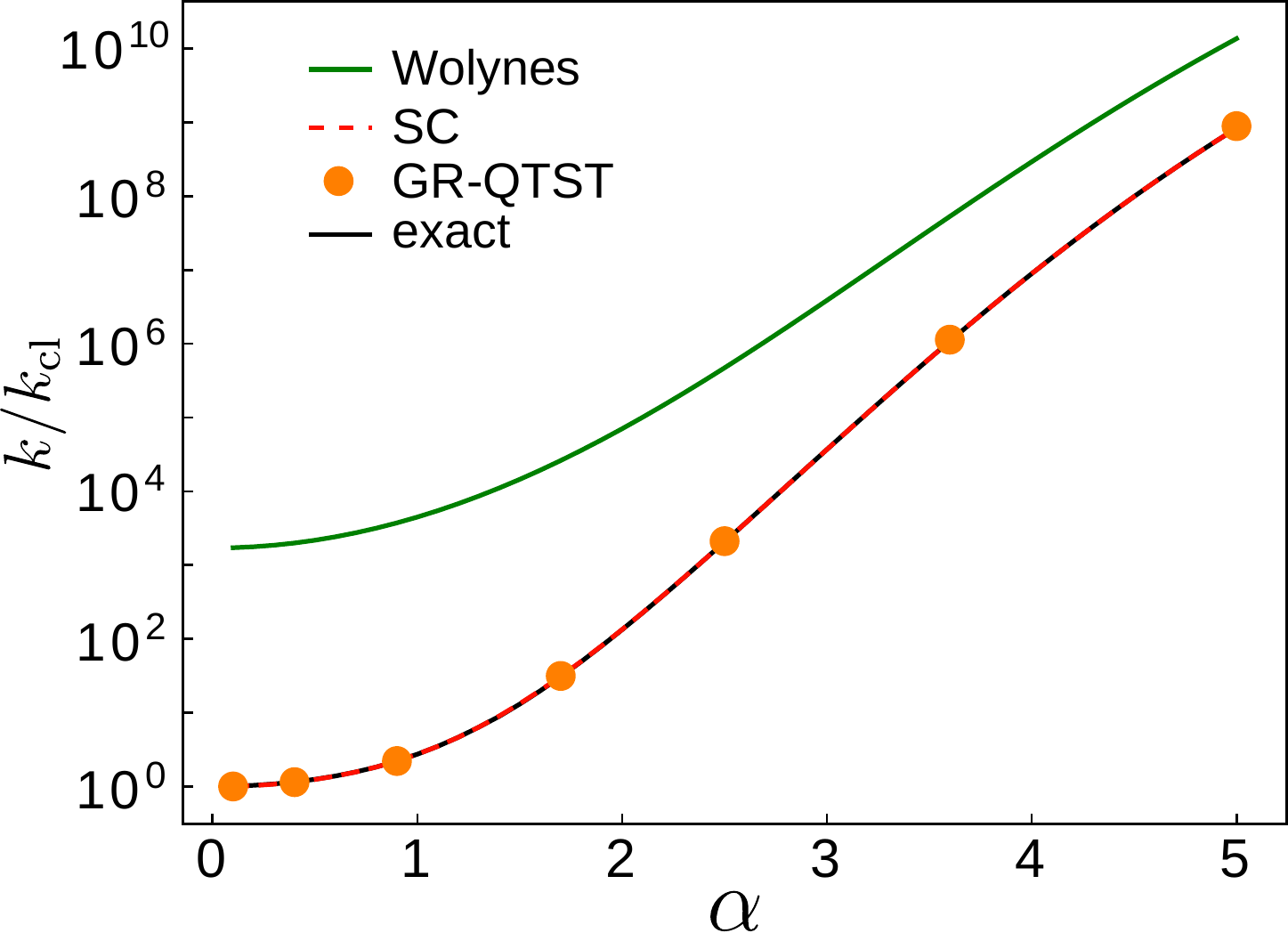}
    \caption{
    Comparison of the Wolynes rate [Eq.~\eqref{wolk}],
	 exact rate [Eq.~\eqref{k}],
	 semiclassical (SC) instanton rate [Eq.~\eqref{instphi}]
	 and GR-QTST rate [Eq.~\eqref{grqtst}]
	 over a range of $\alpha$ values for the 1D model defined in Appendix~\ref{sec:1danalytic}.
    Results are given relative to the classical rate [Eq.~\eqref{LZ}] of the same system.
    }
    \label{fig:bhw}
\end{figure}

Despite the success of the separated version,
standard Wolynes theory clearly overestimates the rates by over two orders of magnitude.
In fact, this problem persists even in the classical limit.
To demonstrate this,
we calculated rates on the 1D system over a range of ``quantumness", defined by parameter $\alpha$, 
as shown in \fig{bhw}.
Small $\alpha$ values correspond to the classical limit, while for large $\alpha$ values nuclear quantum effects are prominent.
We find that the standard Wolynes rate, which is estimated by treating both TSs together,
incorrectly predicts a huge speed-up in the rate relative to a classical calculation, even in the $\alpha\rightarrow0$ limit.
The error in $\log k$ is actually largest in the classical limit for this system
because the two TSs in the model make almost equal contributions to the classical rate,
which causes the $\phi_\text{u}(\tau)$ curve to deviate most strongly from either of the separated effective actions in this limit.
Note however that in the quantum regime, even though the exact rate is completely dominated by transition state A,
the action $\phi_\text{u}(\tau)$ is still significantly perturbed by the presence of the other TS,
leading to a large error in Wolynes theory.

GR-QTST is not based on a saddle-point approximation, but on its connection to instanton theory.
The main difference to Wolynes theory is that 
the rate is obtained from a \emph{constrained} effective action, $\phi_\text{c}(\tau)$, which is defined in \sref{methods} [\eqn{gr_approx}] and includes contributions only from paths with matching reactant and product energies, and which therefore resemble instanton-like configurations.
The GR-QTST method for this one-dimensional system
was evaluated using the same Monte Carlo importance sampling approach employed in Paper I\@.
The results are, at least for low-dimensional systems, approximately independent of $\tau$
and so the exact specification of the value used to evaluate the GR-QTST rate is relatively unimportant.
The ansatz of Paper I used a value of $\tau$ which maximizes the constrained action, $\phi_\text{c}(\tau)$ \cite{GRQTST}.
However, in this paper, we choose to evaluate the rate at the value of $\tau=\tau^*$ which maximizes $\phi_\text{u}(\tau)$.
This is simpler to compute, especially when using the PIMD method described in \sref{methods} and
the slight change of ansatz does not affect any of our conclusions from Paper I\@. 
Although GR-QTST is not rigorously separable for a system with two transition states, it is at least approximately so,
as the rate is approximately independent of $\tau$.
\footnote{Further changes to the GR-QTST ansatz,
such as always choosing $\tau=\beta\hbar/2$ or averaging over a fixed range of values,
would in fact make it rigorously separable in this case.
}
For this reason,
it is seen that GR-QTST accurately predicts the rates for this simple one-dimensional model consisting of two transition states.
This is true whether the method is applied directly to both crossings simultaneously (GR-QTST)
or whether the rate is obtained as the sum of the two individual rates (GR-QTST separated), as can be seen in Table \ref{table:1dtable}\@.
In fact we showed in Paper I that GR-QTST reproduces the exact rate for 
a one-dimensional system of two crossed linear potentials at any temperature.\cite{GRQTST}
Because in this special case, $\phi_\text{c}$ is rigorously independent of $\tau$, it would therefore also be exact for a system of many crossed linear potentials.

Note that there is a danger that for high-dimensional complex systems exhibiting multiple transition states,
GR-QTST in its present form could suffer from similar problems to Wolynes theory,
because it was explained in Paper I that $\phi_\text{c}(\tau)$ becomes increasingly curved as the number of degrees of freedom is increased.
It will therefore be important to address this size-consistency problem when applying it to large systems at low temperatures. 
However, one expects the problem to be less severe than for Wolynes theory because,
unlike Wolynes theory, GR-QTST rigorously tends to the correct classical limit as $\alpha\rightarrow0$,
which is a general property of the method for all systems, irrespective of the number of degrees of freedom and number of transition states.
This is because the GR-QTST method contains a constraint functional which forces the ring polymer (which is collapsed in this limit) to lie on diabatic crossing seam, thereby sampling only configurations at the classical transition state.

\section{Methods}
\label{sec:methods}
Going beyond the simple system discussed in the previous section, it is important to discuss how GR-QTST rate can be computed for electron-transfer reactions in complex models and molecular systems, where the Monte Carlo importance sampling scheme described in Paper I falls short.
Therefore, in this section we introduce
a PIMD implementation of GR-QTST\@. 
In \sref{results}, we will then apply this method to an anharmonic and non-separable two-dimensional system
and demonstrate that GR-QTST alleviates the problems identified with Wolynes and instanton theory.

In Paper I, we presented the formal theory of the GR-QTST method in terms of continuous path integrals
which travel for imaginary-time $\tau$ on the reactant state, $\ket{0}$, and for imaginary-time $\beta\hbar-\tau$ on the product state, $\ket{1}$.
Here we present only the working equations.
As in Wolynes theory \cite{Wolynes1987nonadiabatic},
the path integral is replaced by an ensemble average of discretized imaginary-time paths.
These paths can be represented by a ring polymer of $N\equiv N_0+N_1$ beads,
in which $N_0-1$ beads feel the potential of the reactant state and
$N_1-1$ feel that of the product state.
The remaining two beads are called the ``hopping beads'' and feel a potential which is the average of the two surfaces.
We introduce
$\lambda$ as a dimensionless order parameter 
which determines how the beads are distributed between the two diabatic states
and is defined by
\begin{equation} \label{lambda}
1 - \lambda \equiv \frac{\tau}{\beta\hbar} \equiv \frac{N_0}{N},
\end{equation}
where because $N_0$ and $N$ are integers, only discrete values are allowed.
There are two special cases; when $\lambda=0$, the ring polymer is completely on the reactant electronic state,
and when $\lambda=1$ it is completely on the product electronic state. 

The unconstrained ensemble can be sampled using thermostatted PIMD
\footnote{We employed the standard normal-mode update scheme which we found to be stable.  In case of instability problems in future applications one could instead apply the scheme described in \Ref{Korol2019Cayley}.}\nocite{Korol2019Cayley}
based on the following Hamiltonian:
\begin{subequations}
	\label{HRP2}
\begin{align}
%    H_{\text{RP}}^{(N,N_0)}
    H_{\text{u}}^{(\lambda)} &= \sum_{i=1}^{N}\sum_{j=1}^{D}\frac{[p_j^{(i)}]^2}{2m_j} +U_{\text{RP}}(\mathbf{x}) + U_N^{(\lambda)}(\mathbf{x})\\
    U_\text{RP}(\mathbf{x}) &= \sum_{i=1}^{N}\sum_{j=1}^{D}\frac{1}{2}m_j\omega_N^2[x_j^{(i)}-x_j^{(i-1)}]^2 \\
    U_N^{(\lambda)}(\mathbf{x}) &= \sum_{i=1}^{N_0-1}V_0(\mathsf{x}^{(i)})+\sum_{i=N_0+1}^{N-1}V_1(\mathsf{x}^{(i)})+\sum_{i\in\{N_0,N\}}\frac{1}{2}[V_0(\mathsf{x}^{(i)})+V_1(\mathsf{x}^{(i)})], \quad \text{for $0<\lambda<1$}
	 \\
    U_N^{(0)}(\mathbf{x}) &= \sum_{i=1}^{N}V_0(\mathsf{x}^{(i)}), %\text{if $N_0=N$}
	 \\
    U_N^{(1)}(\mathbf{x}) &= \sum_{i=1}^{N}V_1(\mathsf{x}^{(i)}), %\text{if $N_0=0$}
\end{align}
\end{subequations}
where $\omega_N=1/\beta_N\hbar$, $\beta_N=\beta/N$, $\beta=1/k_\mathrm{B}T$, $k_\mathrm{B}$ is the Boltzmann constant, $T$ is the temperature, 
and $\textbf{x}=\{\mathsf{x}^{(1)},\dots,\mathsf{x}^{(N)}\}$ 
are the positions of the beads with conjugate momenta $\mathbf{p}$.
The cyclic index $i$ runs over each bead such that $\mathsf{x}^{(0)}\equiv \mathsf{x}^{(N)}$, and the index $j$ runs over each 
of the $D$ nuclear degrees of freedom of the system.
The reactant partition function is defined by an ensemble of ring polymers
\begin{align}
    \label{Z0}
	Z_0 = 
	\frac{1}{(2\pi\hbar)^{ND}} \iint \eu{-\beta_N H_{\text{u}}^{(0)}} \, \text{d}\textbf{x} \, \text{d}\textbf{p}.
\end{align}
First we briefly discuss the computation of Wolynes rate theory \cite{Wolynes1987nonadiabatic}, given by:
\begin{equation}
k_{\text{Wolynes}} = \frac{\Delta^2}{\hbar}\sqrt{2\pi\beta}\left(-\frac{\partial^2 F_\text{u}}{\partial\lambda^2}\right)_{\lambda=\lambda^*}^{-\half} \eu{-\beta F_{\text{u}}(\lambda^*)},
\end{equation}
where $F_{\text{u}}(\lambda^*)$ is the maximum unconstrained free energy with respect to $\lambda$
and thus also defines $\tau^*$ through \eqn{lambda}. 
It can be calculated using adiabatic switching and thermodynamic integration \cite{TuckermanBook}
\begin{equation} \label{Fu}
F_{\text{u}}(\lambda) = \int_0^{\lambda} \frac{\partial F_{\text{u}}}{\partial \lambda'} \, \text{d}\lambda',
\end{equation}
where
\begin{equation}
\frac{\partial F_{\text{u}}}{\partial\lambda}=-\langle V_0(\mathsf{x}^{(0)})-V_1(\mathsf{x}^{(0)})\rangle_{\text{u}}^{(\lambda)},
\end{equation}
and $\langle \dots \rangle_{\text{u}}^{(\lambda)}$ denotes the canonical ensemble average with the Hamiltonian in Eq.~(\ref{HRP2}), which can be obtained from PIMD simulations.

The GR-QTST rate formula, as mentioned in previous sections, is defined in terms of a constrained effective action, $\phi_\text{c}(\tau)$, and the reactant partition function, $Z_0$ [\eqn{Z0}].
The constrained action is defined by
\begin{equation}
\eu{-\phi_{\text{c}}(\tau)/\hbar}
= \frac{1}{(2\pi\hbar)^{ND}} \iint \eu{-\beta_N H_{\text{u}}^{(\lambda)}} \, \delta[\sigma_\lambda(\mathbf{x})]
    \, \text{d}\textbf{x} \, \text{d}\textbf{p},
\label{gr_approx}
\end{equation}
and the definition of
the function $\sigma_\lambda(\mathbf{x})$,
which constrains the reactant and product energies to match,
can be found in Appendix~\ref{sec:definitions}\@.
Direct calculation of the constrained action and the partition function individually is difficult using PIMD,
but, as is typical in rate theory, only the
constrained free energy, $F_{\text{c}}(\lambda)$, relative to the reactant %free energy
is actually required.
Although not calculated in this way, this free energy is defined by
$\eu{-\beta F_\text{c}(\lambda)} \equiv \eu{-\phi_\text{c}(\tau)/\hbar} / Z_0$.
Thus the GR-QTST method approximates the golden-rule rate
with \cite{GRQTST}
\begin{equation}
	\label{grqtst}
	k_\text{GR-QTST} = \frac{2\pi\beta\Delta^2}{\hbar} \eu{-\beta F_\text{c}(\lambda^*)}.
\end{equation}
According to the ansatz of Paper I, $\lambda^*$ would be chosen as the value of the order parameter at which the constrained free energy is maximum.
However, in this work, we use the simpler ansatz discussed in the previous section, in which $\lambda^*$ is the same as that used in Wolynes theory, i.e.\ the maximum of the unconstrained free energy.
In order to obtain the constrained free energy,
we first express the delta function as the limit of a function of a new variable $K$,
\begin{equation}
\delta(\sigma_\lambda)=\lim_{K\rightarrow\infty}\sqrt{1+\frac{K}{2\pi}} \, \eu{-\frac{1}{2}K\sigma_\lambda^2}.
\label{del_sig}
\end{equation}
The introduction of this function % $K$ order parameter
makes it feasible to compute the GR-QTST rate with a PIMD free-energy simulation
using the biased Hamiltonian
\begin{equation}
H_{\text{b}}^{(\lambda,K)} = H_{\text{u}}^{(\lambda)} + \frac{1}{2\beta_N}\left[K\sigma_\lambda^2-\ln\left(1+\frac{K}{2\pi}\right)\right]. 
\label{HGR}
\end{equation}
One can see that this Hamiltonian includes an ``umbrella-like" potential of the function $\sigma_\lambda(\mathbf{x})$.
In contrast to the usual umbrella-sampling approaches,\cite{DaanFrenkel_book,TuckermanBook} here the centre of our umbrella is fixed but the strength can be varied.
Note that we have chosen this definition such that when $K=0$,
the unconstrained ensemble will be obtained.
However, as $K$ tends to infinity, it enforces the GR-QTST energy constraint exactly.
\cite{RhiannonThesis}
When performing the PIMD simulations with Hamilton's equations of motion based on $H_{\text{b}}^{(\lambda,K)}$,
an extra term appears in the first derivative due to the umbrella:
\begin{equation}
\frac{\partial H_{\text{b}}^{(\lambda,K)}}{\partial \textbf{x}} = \frac{\partial H_{\text{u}}^{(\lambda)}}{\partial \textbf{x}} + \frac{K\sigma_\lambda}{\beta_N}\frac{\partial \sigma_\lambda}{\partial \textbf{x}}.
\end{equation}
An explicit formula for the derivative of $\sigma_\lambda$
is given in Appendix~\ref{sec:definitions}\@.
Given that the unconstrained free energy $F_{\text{u}}(\lambda)$ has been calculated as described above [\eqn{Fu}],
we propose the following scheme for calculating the constrained free energy, 
\cite{RhiannonThesis}
\begin{equation}
F_{\text{c}}(\lambda) = F_{\text{u}}(\lambda) + \int_0^{\infty}\frac{\partial F_{\text{b}}(\lambda,K)}{\partial K} \, \text{d}K,
\label{F_TI}
\end{equation}
where
\begin{equation}
\frac{\partial F_{\text{b}}(\lambda,K)}{\partial K} = \frac{1}{2\beta}\left[\langle\sigma_\lambda^2\rangle_{\text{b}}^{(\lambda,K)}-\frac{1}{K+2\pi}\right],
\end{equation}
and $\langle \dots \rangle_{\text{b}}^{(\lambda,K)}$ denotes the canonical ensemble average with the biased Hamiltonian in Eq.~\eqref{HGR}.
Here we are using thermodynamic integration (TI)
to measure the free-energy change upon slowing turning on the constraint
and 
we will call this method for evaluating constrained free energies ``$\delta$-TI''.

To avoid integration to infinity in Eq.~\eqref{F_TI}
and to greatly improve numerical stability,
the integral over $K$ is performed
using the coordinate transform
\begin{equation} \label{TIcoord}
\xi = 1-\frac{1}{1+\sqrt{K/K_0}},
\end{equation}
where $\xi$ ranges from 0 to 1 and $K_0$ is a scaling parameter, which we chose to be 0.1 in the following.
The integral can now be performed using
\begin{equation} \label{F_TI_xi}
%\int_0^{\infty}\frac{\partial F_{\text{b}}(\lambda,K)}{\partial K}\text{d}K = 
F_{\text{c}}(\lambda) = F_{\text{u}}(\lambda) + 
\int_0^{1}\frac{\partial F_{\text{b}}(\lambda,K)}{\partial K}\frac{\text{d} K}{\text{d}\xi} \, \text{d}\xi.
\end{equation}
Tests showed that the results barely depend on $K_0$ values over two orders of magnitude, $K_0\in[0.01,1]$.
We note that there are also other coordinate transformations that one could employ to perform the integral,
but it was found that the above definition yields results reliably and efficiently for the systems studied here.

With the new ansatz, only one $\delta$-TI needs to be performed to obtain the GR-QTST rate, which is for $\lambda=\lambda^*$.
This is simpler to compute compared with the GR-QTST ansatz of Paper I, for which several $\delta$-TI calculations at different values of $\lambda$ would be required.
Our results will show that $F_\text{c}$ is approximately independent of $\lambda$ for the systems investigated in this work,
and thus the results are not significantly affected by this change.

Note that one could conceive of applying other free-energy sampling schemes to obtain the constrained free energy,
which may have computational advantages.
Possible alternative approaches which could be employed in future work
include thermodynamic integration of the potential of mean force obtained from a constrained PIMD, metadynamics or adiabatic free-energy dynamics \cite{TuckermanBook},
each using $\sigma_\lambda$ as the reaction coordinate. 

The classical rate in the golden-rule limit is defined by \cite{nonoscillatory}
\begin{equation}
k_{\text{cl}}
    = \frac{2\pi\beta\Delta^2}{\hbar} \, \eu{-\beta F_{\text{cl}}},
\label{kcl}
\end{equation}
where
\begin{equation}
    Z_0 \, \eu{-\beta F_{\text{cl}}}
    \equiv \frac{1}{(2\pi\hbar)^{ND}} \iint \eu{-\beta H_0(\textbf{x})}\delta[\beta(V_0(\textbf{x})-V_1(\textbf{x}))] \,\text{d}\textbf{x}\,\text{d}\textbf{p}.
\end{equation}
Here
$F_{\text{cl}}$ is the classical free energy 
of the crossing seam relative to 
the reactant well.
It can be defined in a similar way to the ring-polymer versions given above with $N=1$ and $\lambda=0$ % N_0=1
and can also be calculated with a $\delta$-TI approach similar to
\eqn{F_TI_xi}.
Finally, we shall also employ
the semiclassical instanton method as described in \Ref{GoldenRPI}
and the steepest-descent approximation (harmonic limit) to the classical rate, which can be calculated with the procedure described in \Ref{nonoscillatory}.

\section{Results}
\label{sec:results}

In the model system described in \sref{theory},
we were able to avoid the problems associated with Wolynes theory
by computing the rate separately for the two product channels.
However, realistic systems are complex and multidimensional,
and typically more than one transition state may exist for each product channel.
Therefore performing Wolynes theory for each transition state separately will not be possible.
In this section, we introduce a 2D model system in which there are two transition states, but only one product state.
We show that
Wolynes theory breaks down in the same way as shown in the 1D system
but that nonetheless
the GR-QTST method continues to give correct order-of-magnitude estimates for the rate constant when compared with numerically exact results.

Using the coordinates $\mathsf{x}=(x,y)$, the potentials of our 2D model are defined as
\begin{subequations} \label{pot2D}
\begin{align}
V_0(x,y)&=h(y)[x+x_0(y)]^2 + h_0\exp(-a_0 x) + V_{\text{r}}(y)\\
V_1(x,y)&=h_1\exp(-a_1 x) + V_{\text{r}}(y),
\end{align}
where
\begin{align}
h(y) &= p_1(1+\tanh(-y/d))\\
x_0(y) &= p_2\exp(p_3y)\\
V_{\text{r}}(y) &= h_2y^2+h_6y^6.
\end{align}
\end{subequations}
$V_1$ is chosen to be a separable system
so as to simplify the computation of the exact rate. 
However, the overall system is coupled and thus provides a nontrivial test for the methods.

The following set of parameters defines ``system I'': $h_0=4.8$, $a_0=0.4$, $h_1=32$, $a_1=1$, $h_2=3.2$, $h_6=0.32$, $d=0.75$, $p_1=28.8$, $p_2=1.15$, $p_3=0.7$ (using reduced units with $\hbar=1$).
This system was designed to ensure that the gradient of $V_0$ is very different
at the two transition states (they differ by a factor of 6.5),
which means that the optimal $\tau$ values for the two crossings will be far apart.
In \sref{anharmonic}, we will tune some of the these parameters to create ``system II".

\begin{figure}[!ht]
    \centering
    \includegraphics[width=12.5cm]{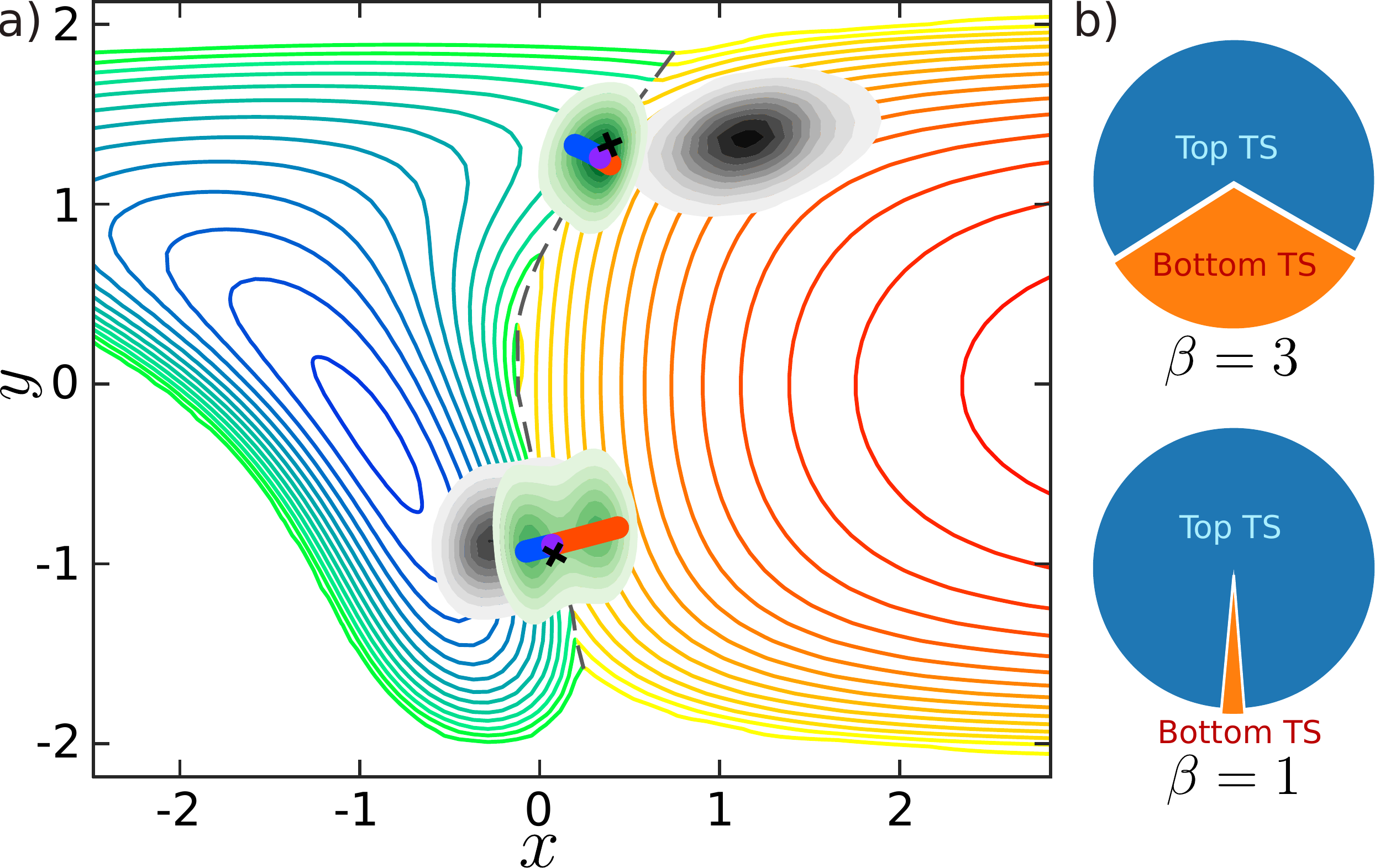}
    \caption{
    (a) Contour plot of the two potential-energy surfaces of system I\@. 
    The blue-green contours show $V_0$ and the red-yellow contours show $V_1$. 
	Only the lower of the two potential-energy surfaces is shown at each coordinate and 
    the dashed line shows the crossing seam between the two surfaces.
	The `$\times$'s mark the two classical transition states on the crossing seam and
    the two instantons for $\beta=3$ are also shown, with blue beads on $V_0$, red on on $V_1$, whereas the violet bead indicates the hopping point.
    The filled contours represent density plots of the bead positions at $\beta=3$ for unconstrained/Wolynes (grey) and constrained/GR-QTST (green) obtained from simulations with $\lambda=\lambda^*$.
    (b) A representation of the ``mechanistic branching fraction,'' defined in terms of contributions to the total rate through the two TSs, here according to instanton theory at $\beta=3$ and $\beta=1$.
    }
    \label{pot1}
\end{figure}
A plot of the PES of system I is shown in Fig.~\ref{pot1}(a).
The potential mimics complex reactions in solution,
where more than one TS exists due to the fluxional anharmonic solvent environment.
We will study the electron-transfer rate for a particle with mass 35 
at low temperature $\beta=3$.
Although these parameters were chosen to cause problems for Wolynes theory,
we note that the breakdown of Wolynes theory occurs more generally, as was explained in \sref{theory}. 
To demonstrate this numerically,
we also studied this system at higher temperature $\beta=1$
where the reaction behaves classically.

\subsection{Computational details}
We computed rates in the golden-rule limit for system I using 
classical golden-rule transition-state theory, \cite{nonoscillatory,ChandlerET}
Wolynes theory, \cite{Wolynes1987nonadiabatic}
semiclassical instanton theory, \cite{GoldenGreens,GoldenRPI}
and the newly developed GR-QTST method. \cite{GRQTST}
Instanton theory was applied 
using the ring-polymer approach
based on the Lagrangian formalism described in \Ref{GoldenRPI}
with 64 beads on each PES.
In this method, the instanton configurations are obtained
by a first-order saddle-point optimization in the combined space of $\textbf{x}$ (bead positions) and $\tau$ variables.
For comparison, numerically exact quantum results were calculated using a discrete variable representation (DVR).\cite{Meyer1970DVR,Light1985DVR}
A grid of $200\times50$~DVR points was used, %with $\omega=10/m$ was used (\red{define carefully omega}) for $V_0$, 
and we tested that a larger grid of $250\times80$~DVR points gives the same results.
$V_1$ is separable in the $x$ and $y$ degrees of freedom, hence we used the same one-dimensional DVR as for $V_0$ in the $y$ degree of freedom, and analytical scattering wave functions for the $x$ degree of freedom \cite{Bunkin1973Morse}.
We performed Wolynes rate calculations with PIMD simulations using
the Andersen thermostat 
and $N=128$ ring-polymer beads to represent the imaginary-time paths.
At $\beta=3$, each PIMD simulation ran for 108,000 steps with a time-step of 0.03, with the first 8,000 discarded for equilibriation.
For $56 \le N_0 \le 72$, the simulations ran for 208,000 steps and employed
replica exchange with 36 replicas (with the ``hottest" replica at $\beta=0.105$) to ensure ergodic sampling of the two crossings (see Appendix~\ref{sec:replica} for details).
At $\beta=1$, the simulation setup was almost the same as at $\beta=3$, with the only difference in the replica-exchange setup.
Here, replica-exchange sampling was performed only for $68 \le N_0 \le 96$, with 16 replicas (the ``hottest" replica at $\beta=0.11$) and 308,000 simulation steps.
The computed free-energy integrands are shown in the supporting information (SI).%
\footnote{Supplementary material at xxxxx}
The GR-QTST simulations were also performed using $N=128$. 
The simulation setup was mostly the same as the Wolynes simulations (108,000 steps), with two differences.
First, a shorter time-step of 0.005 was used for simulations with $K\ge1$, because the constraining forces increase with the value of $K$.
This does mean that the simulations become less efficient at large $K$, but nonetheless it was not a problem to converge the results for the systems investigated, such that we could obtain reasonably small error bars. 
Second, replica-exchange sampling was applied for all simulations, whatever the value of $N_0$.
The thermodynamic integration used at least 12 different $K$ values in the range $[0.0005,16]$.
The integral was computed with Simpson's rule, although
we noted that using the trapezoidal rule instead gives the values within or comparable to the error bars of the results.
The classical rates 
were calculated with
molecular dynamics simulations which ran for 205,000 steps with the first 5,000 discarded for equilibriation, at 17 different $K$ values in the range [0.0001,16].
The time-step and replica-exchange setup were the same as in the GR-QTST simulations.
The computational setup for system II was almost the same as for system I for all the methods, with the only difference in the replica-exchange setup.
The Wolynes calculations on system II used replica exchange with 16 replicas (with the ``hottest" replica at $\beta=0.15$) for $42 \le N_0 \le 60$, and 8 replicas for $60 \le N_0 \le 70$, which ran for 208,000 steps.
The GR-QTST calculation used 16 replicas with the ``hottest" replica at $\beta=0.15$, which ran for 108,000 steps.
The free-energy integrands are shown in the SI.

\subsection{Benchmarking the rates calculated using various methods}

\begin{table}[!ht]
\caption{
Comparison of the golden-rule rates, $k/\Delta^2$, (with powers of 10 in parentheses) in reduced units for systems I and II, calculated with various methods.
The classical rate is computed in two different ways: once in full using molecular dynamics (MD) and once using the steepest-descent (SD) approximation, which is equivalent to a local harmonic analysis around the TS\@.
The standard error in each simulation is estimated in the standard way taking into account the autocorrelation time \cite{DaanFrenkel_book}.
}
\label{ratecmp_p1}
\begin{tabular}{l@{\extracolsep{6pt}}rrr@{}}
\hline
\hline
               & \multicolumn{2}{c}{System I}                  & \multicolumn{1}{c}{System II}                  \\
\cline{2-3}\cline{4-4}
         & \multicolumn{1}{c}{$\beta=3$} & \multicolumn{1}{c}{$\beta=1$} & \multicolumn{1}{c}{$\beta=3$} \\
\hline
classical (SD) & 1.39(-29)~~~                           & 9.42(-11)~~~                        & 2.82(-26)      \\
classical (MD) & 1.1$\pm$0.2(-29)~~~                    & 7.3$\pm$0.7(-11)~~~                 & 2.7$\pm$0.2(-26)     \\
instanton   & 2.07(-28)~~~                           & 1.11(-10)~~~                        & 4.28(-23)      \\
Wolynes        & 6.6$\pm$0.4(-26)~~~                    & 2.69$\pm$0.05(-10)~~~               & 1.23$\pm$0.06(-23)     \\
% GR-QTST        & 1.6$\pm$0.2(-28)~~~                    & 1.02$\pm$0.07(-10)~~~               & 6.7$\pm$0.7(-24)      \\
GR-QTST        & 2.3$\pm$0.2(-28)~~~                    & 1.02$\pm$0.07(-10)~~~               & 7.6$\pm$0.6(-24)      \\
exact          & 1.98(-28)~~~                           & 1.07(-10)~~~                        & 7.10(-24)          \\
\hline
\hline
\end{tabular}
\end{table}
Here we discuss only the results obtained for system I, which are summarized in Table~\ref{ratecmp_p1},
and leave the discussion of system II to \sref{anharmonic}.
There are two transition states (which we call ``top'' and ``bottom'') each with its own instanton, as shown in Fig.~\ref{pot1}(a).
The two transition states have a different character such that (at $\beta=3$)
the top instanton corresponds to $\lambda=0.29$ whereas the bottom instanton corresponds to $\lambda=0.72$.
Instanton theory predicts that at this temperature,
the reaction rate through either instanton is roughly equal, as shown in Fig.~\ref{pot1}(b).
This occurs because although the bottom transition state is higher in energy, it has a thinner barrier which is more conducive to tunneling.

\begin{figure}
    \centering
    \includegraphics[width=14cm]{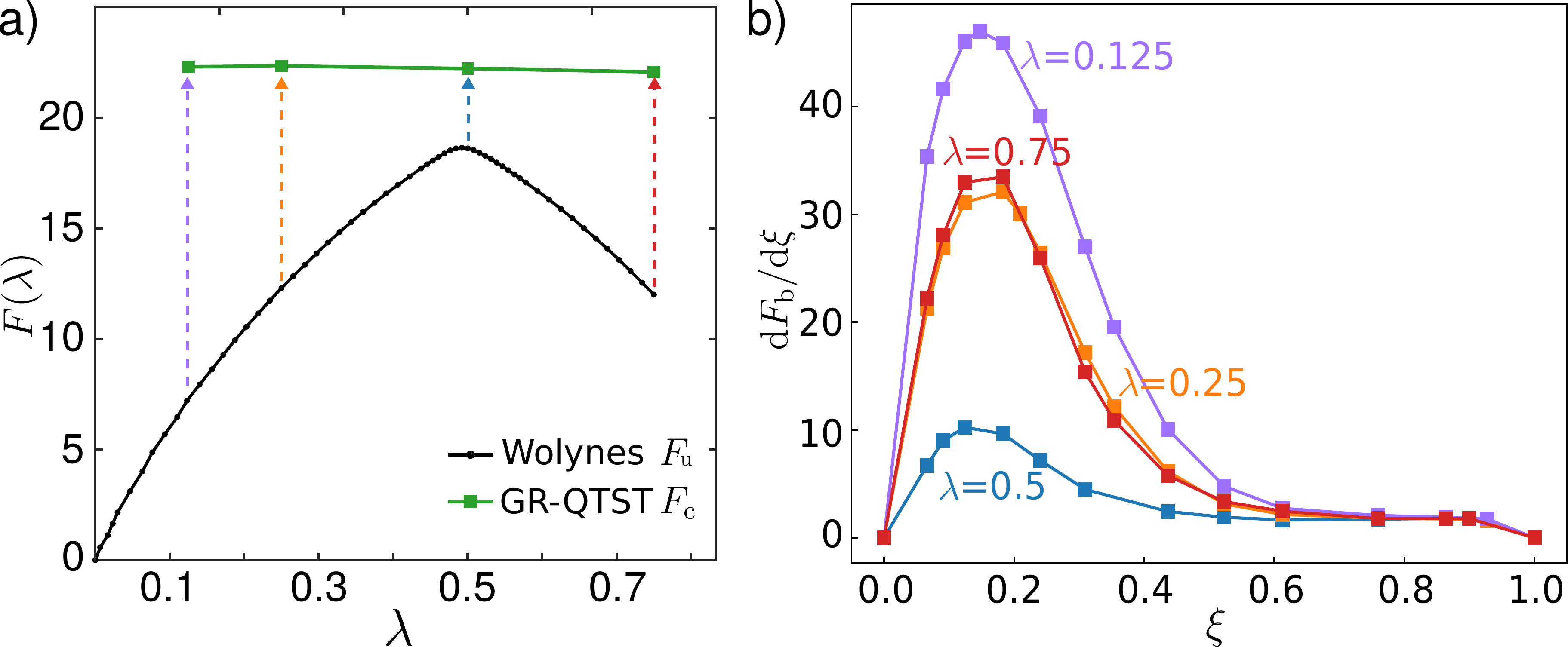}
    \caption{
	 Free-energy calculations.
    (a) Comparison of the $\lambda$ dependence of the unconstrained free energy (used for Wolynes theory) and the constrained free energy (used for GR-QTST) for system I at $\beta=3$.
    The dashed arrows illustrate how the GR-QTST free energies are obtained (Eq.~\eqref{F_TI}).
    (b) The constrained free-energy integrand at different $\lambda$ values obtained from biased PIMD simulations.
    The colours correspond to the colours of the dashed arrows in (a).
    The statistical error bars in this case are smaller than the symbol size.
    }
    \label{FGRQTST}
\end{figure}

We find that the steepest-descent approximation does not lead to significant errors for this system.
This is the case both for the harmonic classical rate, which is fairly close to the full classical result,
as well as for the instanton, which is close to the exact quantum result.
However, at $\beta=3$, both classical methods underestimate the rate by more than an order of magnitude
because nuclear quantum effects such as tunneling are not accounted for.
Instanton theory is thus seen to describe this tunneling effect very accurately,
as expected from previous work \cite{AsymSysBath},
and shows good agreement with the exact results,
with only a 5\% error mostly emanating from the harmonic approximation to the reactant partition function.
Wolynes theory however, overestimates the rate by two orders of magnitude,
and actually $\log k$ has a larger error (albeit in the opposite direction) than the classical rate for this system.
This example shows that the break-down of Wolynes theory identified in \sref{theory}
is a general phenomenon, which may exist in many realistic reactions.
The GR-QTST rates are calculated using the $\delta$-TI procedure described in \sref{methods},
and we show an example of the constrained free energy obtained and the thermodynamic integration over $\xi$ in Fig.~\ref{FGRQTST} (similar plots are given in the SI for the other systems).
One can see that under the GR-QTST energy constraint, the free energy becomes almost independent of $\lambda$,
implying that the GR-QTST rate will not be strongly affected by our choice of ansatz.
The free-energy integrands show that the largest effect of the constraint comes from $\xi<0.5$, corresponding to small $K$ values ($K<0.1$).
The GR-QTST rate agrees well with the exact result, with only $\sim$15\% error at $\beta=3$ (which is comparable to the statistical uncertainty of the result).
This confirms that the energy constraint cures the problem of Wolynes theory in these systems.

At a higher temperature ($\beta=1$), quantum tunneling becomes less important and the classical rates do not differ much from the exact quantum rate.
Instanton theory also performs well as it is known that in this limit it tends to the harmonic classical rate,
\cite{GoldenGreens}
which is a good approximation in this case.
GR-QTST also rigorously tends to the correct classical limit, \cite{GRQTST}
predicting a rate within statistical error of the exact result.
The Wolynes rate, on the other hand, does not tend to the correct limit, and it suffers from the same problems for this system as it did in the $\beta=3$ case, overestimating the rate by a factor of 2.5 here. 

These finding indicate that Wolynes theory could drastically overestimate the rate,
seemingly predicting huge tunneling effects in systems that have multiple transition states, even in the high-temperature limit where a classical approximation would be valid.
However, by including an energy constraint in the path-integral sampling, GR-QTST is able to give reliable predictions for the rate,
both at high and low temperatures,
with a similar accuracy to the rigorously-derived instanton theory.

\subsection{Reaction mechanism}
Obtaining an accurate prediction of the rate is only one objective for simulations of electron-transfer reactions, and
characterizing the correct reaction mechanism is arguably even more important. % than obtaining an accurate reaction rate.
Here we compare the mechanistic information obtained from Wolynes theory and from GR-QTST,
based on the ensemble of ring polymers sampled by these methods.
These should correspond to a transition-state ensemble and thus the relative populations at two transition states should quantify the 
``mechanistic branching fraction'', which we will define as probability for a reactive particle to use one of the two available pathways.
We use the two instanton configurations 
and the values of two separate semiclassical instanton rate calculations to give a benchmark, i.e.\ $k_\text{SC}^{(s)}/k_\text{SC}$, where here $s$ labels either the top or bottom TS.

The unconstrained ensemble described in \sref{methods} with $\lambda=\lambda^*$ is used for a mechanistic analysis of Wolynes theory.
On the other hand, the relevant ring-polymer ensemble for GR-QTST is obtained by
performing constrained PIMD simulations (also with $\lambda=\lambda^*$) with the constraint $\sigma=0$, using the RATTLE algorithm \cite{RATTLE}, and re-weighting to recover the soft constraint limit \cite{DaanFrenkel_book,TuckermanBook}.
Assuming that the ensemble makes a clear split into two separate regions of configuration space, the probabilities corresponding to the mechanistic branching fraction can be extracted.

\begin{figure}[!ht]
    \centering
    \includegraphics[width=10cm]{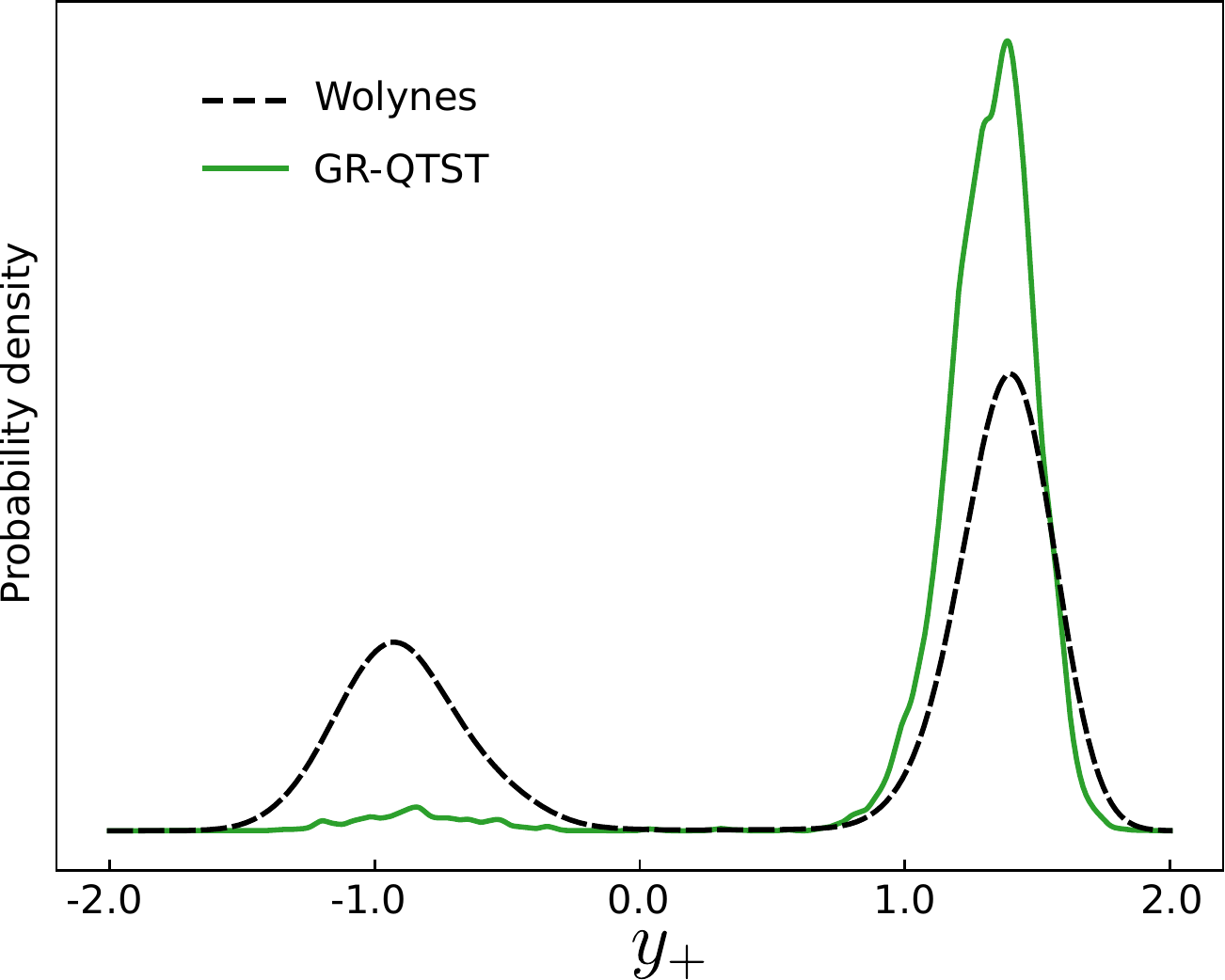}
    \caption{
    Comparison of the distribution of $y_+$ variables, defined as the $y$-component of the average hopping point (see Appendix~\ref{sec:definitions}),
    sampled by the unconstrained ensemble of Wolynes theory and the constrained ensemble of GR-QTST, both at $\lambda=\lambda^*$ with $\beta=1$.
    }
    \label{cmp_xpts_p1b1}
\end{figure}

The density plots of the ensembles sampled from constrained and unconstrained simulations %at their optimal values of $\lambda$
are shown in Fig.~\ref{pot1}(a).
One sees immediately that the transition-state ensemble according to Wolynes theory
does not include the top instanton path and will therefore give an unphysical interpretation of the mechanism of this reaction.
\footnote{In fact, we show in the SI that the unconstrained ensemble remains far from either instanton regardless of the $\lambda$ value used.}
On the other hand, in GR-QTST, the energy constraint ensures that the samples remain close to the crossing seam and thus closer to the instantons in the system, predicting the correct reaction pathways.
The branching fraction for the mechanism through the bottom TS is given by instanton theory as 32\%.
A similar result of 39\% is found by GR-QTST and 44\% by Wolynes theory.
Therefore, despite the fact that Wolynes theory overpredicts the rate by two orders of magnitude, 
the prediction for the branching fraction is seemingly not so bad.
As we will explain shortly, this is a coincidence due to the fact that the two pathways have a roughly equal contribution.
In the higher-temperature case ($\beta=1$) where the Wolynes theory predicts a more reasonable reaction rate (within an order-of-magnitude of the exact result), the predicted reaction mechanism is, however, qualitatively incorrect.
As shown in Fig.~\ref{pot1}(b),
instanton theory predicts that at this temperature transitions through the top transition state is the major contribution to the rate,
whilst only 3\% of the reactive particles will pass through the higher-energy bottom transition state.
As shown in Fig.~\ref{cmp_xpts_p1b1}, Wolynes theory yields an incorrect physical picture, in which both TSs have a significant contribution to the rate at $\beta=1$, with the bottom TS contributing 36\% to the reaction. 
In contrast, GR-QTST identifies that only the top TS is important, and that the bottom TS has only a small contribution of 3\%, giving the correct physical picture, in agreement with the rigorously-derived instanton approach. 

In fact, whenever Wolynes theory breaks down for systems with two TSs,
it will typically sample both TSs as if they are equally important pathways
and give a mechanistic branching fraction of roughly 50\%, regardless of the true dynamics.
We can justify this statement based on Fig.~\ref{phiplot} and the discussion in \sref{theory}
which suggest that $\tau^*$ will typically be found approximately at the location where the two constituent $\phi_\text{u}^{(s)}(\tau)$ terms are equal to each other
and thus give roughly equal contributions to the rate.

\subsection{System with strong anharmonic fluctuations}
\label{sec:anharmonic}

The GR-QTST method
was developed to go beyond the semiclassical approximation of instanton theory.
The accuracy of the semiclassical approximation depends on the anharmonicity of the fluctuations around the instanton path.
In this section, we design a strongly anharmonic system,
for which instanton theory leads to a significant error,
and test the accuracy of the GR-QTST against exact benchmark results.

\begin{figure}
    \centering
    \includegraphics[width=10cm]{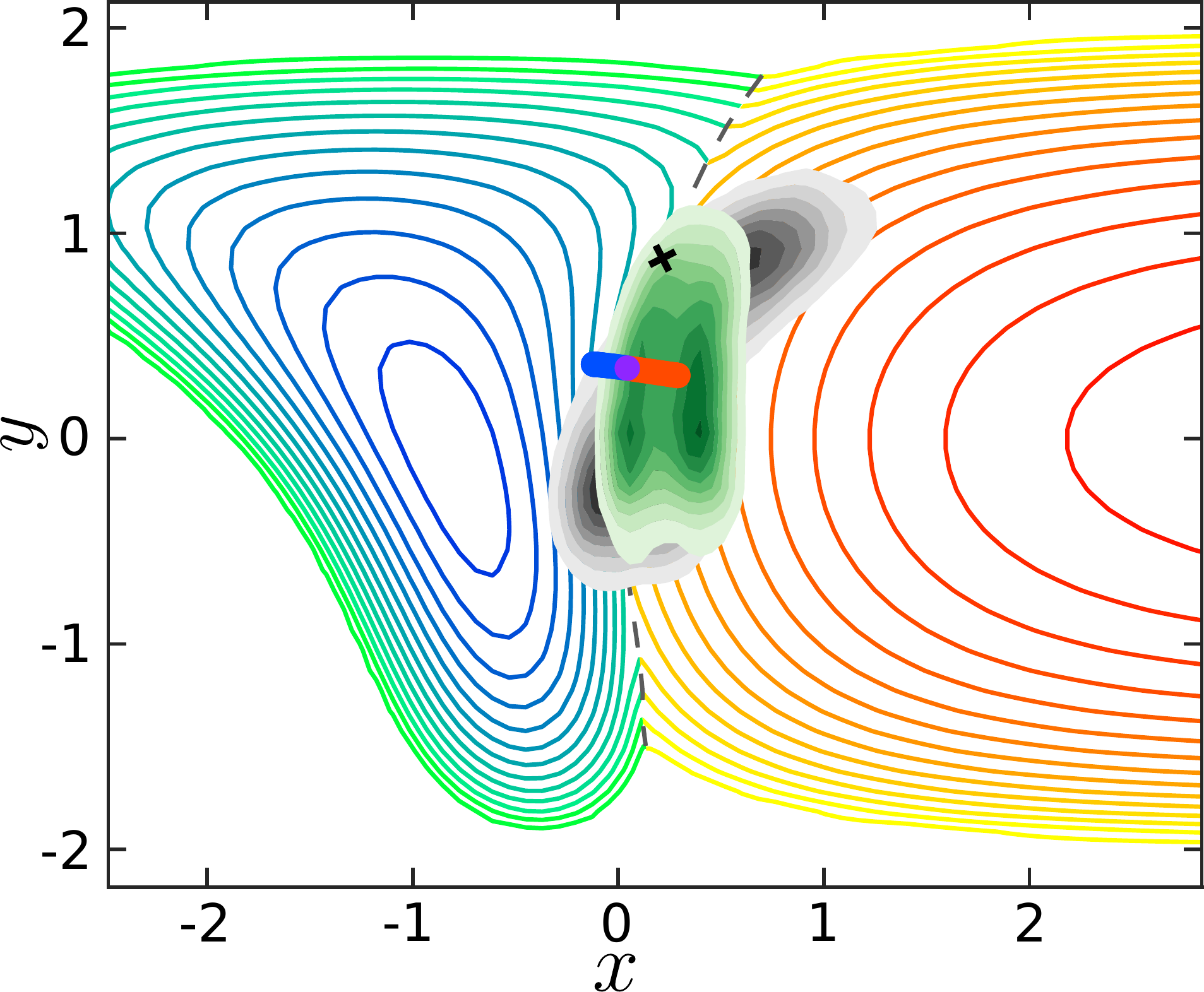}
    \caption{
    Contour plot (as in Fig.~\ref{pot1}) of system II, which has only one TS.
    Here only one instanton exists at $\beta=3$. %, located far from the classical saddle point
    Due to the strong anharmonicity
	the instanton has clearly moved far from the TS due to corner-cutting effects.
    The filled contours show density plots of the bead positions at $\beta=3$ for unconstrained/Wolynes (grey) and constrained/GR-QTST (green) ensembles obtained from simulations with $\lambda=\lambda^*$.
    }
    \label{pot2}
\end{figure}

We modified some of the parameters of system I to create system II\@.
In particular, we defined $d=0.9$ and then chose new values of three parameters $h_2=4.848$, $p_2=0.7145$, $p_3=0.59$ to give an almost zero-frequency normal mode of the instanton at the temperature $\beta=3$.
\footnote{Note that the instanton orbits of golden-rule rate theory [\Ref{GoldenRPI}] do not have a zero-frequency permutational mode like the instantons for adiabatic reactions [\Ref{RPInst}]}
The existence of this very low frequency causes higher-order fluctuations to dominate the path integral,
which are neglected by the semiclassical approximation.
A contour plot of system II together with the instanton is shown in Fig.~\ref{pot2}.
We note that, with system II in particular, it would have been very inefficient
to use the importance sampling approach based on normal modes as described in our previous work \cite{GRQTST}.
However, the new PIMD implementation allows such anharmonic systems to be simulated with a reasonable computational effort.

The golden-rule rate constants calculated with various methods are given in the final column of Table~\ref{ratecmp_p1}\@.
The classical rates are more than two orders of magnitude lower than the exact benchmark,
suggesting that this reaction is in the deep tunneling regime where nuclear quantum effects are prominent.
Note that there is close agreement between the classical harmonic approximation and the full classical rate calculated with molecular dynamics sampling.
This is because
the classical transition state does not have frequencies close to zero
and thus anharmonic effects are not so important here.
Instanton theory, on the other hand, overpredicts the rate by almost an order of magnitude.
This is because 
the neglect of higher-order fluctuations causes it 
to overestimate the true impact of fluctuations around the path. 
The Wolynes rate is $\sim$70\% too high, because it suffers from the same problems 
discussed in the previous section, albeit to a lesser extent for this set of parameters. 
The GR-QTST method makes neither the saddle-point approximation nor the semiclassical approximation,
and because of this, is seen to predict a rate  
much closer to the exact results.

\section{Conclusions}
\label{sec:conclusions}
In this work, we have presented an approach for computing the GR-QTST rate from PIMD simulations,
allowing the method to be applied efficiently to multidimensional anharmonic systems.
We have applied this method to a model system with two transition states 
and demonstrated that GR-QTST gives accurate predictions for the golden-rule rate constant,
whereas Wolynes theory can drastically overestimate the rate. 
This error of Wolynes theory can occur even in the high-temperature regime,
such that there is a danger of predicting strong quantum tunneling effects where there are none. 
We explained that the cause of this problem is the saddle-point approximation
employed by Wolynes theory.
The energy constraint in the GR-QTST method ensures that paths are sampled near to the instantons of this system,
whereas Wolynes theory does not, and it cannot therefore be used to extract mechanistic information.
The GR-QTST method also predicts accurate rates for a system where semiclassical instanton theory breaks down due to its neglect of anharmonicity in the fluctuations around the instanton.
Together, these promising results indicate that, at least for certain types of systems, the GR-QTST method 
is the most accurate imaginary-time path-integral approach for calculating electron-transfer rates in the golden-rule limit.
The GR-QTST method 
does however still require further development and testing.
In Paper I \cite{GRQTST}, we explained that the method, as it is currently defined, is not rigorously size consistent in general,
although it nonetheless was found to give accurate results for a multidimensional spin-boson model
and can be shown to rigorously give the correct result in the classical limit for any system.
In the future, we will test whether GR-QTST can maintain this good performance when applied to 
atomistic simulations of solvated reactions.

\section*{Acknowledgements}
We would like to thank Rhiannon Zarotiadis, who tested some early ideas which led to the results presented in this paper,
as well as Joe Lawrence and Philippe H\"unenberger for useful discussions.
The authors acknowledge financial support from the Swiss National Science Foundation
through Project 175696. 
M.J.T. is grateful for the ETH Zurich Research Grant which supports his doctoral studies.

\appendix
\section{Analytic results for 1D system}
\label{sec:1danalytic}
In this section, we present closed-form expressions for the rate defined by various methods
for the one-dimensional potential-energy surfaces given by
\begin{align}
\label{AnharmonicSys}
\begin{split}
&V_0(x)=\tfrac{1}{2} m \omega^2 x^2
\\
&V_{1}^{(s)}(x)=\kappa_s (x-x_0) + V_0(x_0),
\end{split}
\end{align}
where $m$ is the mass and $\omega$ is the frequency of the reactant harmonic oscillator.
There are two product states labelled by the index $s\in\{\text{A},\text{B}\}$,
where $\kappa_{s}$ is the slope of their linear surfaces and $x_0$ is the location of both transition states, which have the same energy by construction.

We perform calculations in reduced-dimensional units.
The degree of the quantumness is defined by $\alpha=\beta\hbar\omega$, the dimensionless barrier height by $\Phi=\beta m \omega^2 x_0^2/2$ and the reduced gradient by $\eta_{s}=-\frac{\kappa_{s}}{m \omega^2 x_0}$.
At first, we study the system in the quantum regime where tunneling is significant.
We have chosen $\alpha=2.5$, $\Phi=45$, $\eta_\text{A}=0.5$ and $\eta_\text{B}=2$ to obtain the results in Table \ref{table:1dtable}\@.
Results are presented as the dimensionless ratio $k/k_{\text{cl}}$, where $k_\text{cl}=k_\text{cl}^{(\text{A})}+k_\text{cl}^{(\text{B})}$. 

The path integral representation of \eqn{exact}
is Gaussian and can be evaluated analytically such that 
the exact effective action for the system of \eqn{AnharmonicSys} can be written in dimensionless units as follows:
\begin{align}
\label{phi1d}
\eu{-\phi_\text{u}^{(s)}(\tau)/\hbar} = Z^\ddag(\tau) \, \eu{-\tilde{S}^{(s)}(\tau)/\hbar},
\end{align}
where the classical action $\tilde{S}^{(s)}(\tau)$ is
\begin{align} \label{Su}
     \tilde{S}^{(s)}(\tau)=\Phi\hbar \bigg(1-\frac{\tau}{\beta\hbar}\bigg)\Bigg[1+2 \eta_{s}-\frac{1}{12}\eta_{s}^2 \alpha^2\bigg(1-\frac{\tau}{\beta\hbar}\bigg)^2-\frac{1}{2} \eta_{s}^2 \alpha \bigg(1-\frac{\tau}{\beta\hbar}\bigg) \coth \bigg(\frac{\alpha \tau}{2 \beta\hbar}\bigg)\Bigg]
\end{align}
and the effective transition-state partition function is
\begin{align}
	Z^\ddag(\tau)=\sqrt{\frac{\csch (\frac{\alpha \tau}{\beta\hbar})}{\alpha(1-\frac{\tau}{\beta\hbar})+2 \tanh (\frac{\alpha \tau}{2 \beta\hbar})}}.
\end{align}
These formulae can be used in \eqn{k} with complex argument, $\tau-\iu t$, to obtain the exact result by numerical integration over time.
Alternatively, they can be used with $t=0$ to evaluate the result of Wolynes theory, \eqn{wolk}, by numerically solving for the optimal $\tau^*$.
The reactant partition function in reduced dimensions is simply $Z_0=[2 \sinh (\frac{1}{2}\alpha)]^{-1}$ and is not affected by the semiclassical approximation.

The semiclassical approximation is very accurate for this system,
and the following formula was used to compute the instanton rate \cite{GoldenGreens}:
\begin{align}
\label{instphi}
	k_{\textrm{SC}}^{(s)} Z_0
	&= \frac{\Delta^2}{\hbar^2} \sqrt{2 \pi \hbar}
	\bigg[-\frac{\textrm{d} ^2 \tilde{S}^{(s)}}{\textrm{d} \tau^2}\bigg]_{\tau=\tilde{\tau}_s}^{-\frac{1}{2}}
	Z^\ddag(\tilde\tau_s) \eu{-\tilde{S}^{(s)}(\tilde{\tau}_s)/\hbar}.
\end{align}
This expression is evaluated at $\tau=\tilde{\tau}_s$ which satisfies the equality $\frac{\textrm{d} \tilde{S}^{(s)}}{\textrm{d} \tau}=0$.
We can study the expression in certain limiting cases similarly to what was done with the instanton rate for the Eckart barrier \cite{Faraday,InstReview}.
In particular, the semiclassical formula becomes exact in the limit $\hbar\rightarrow0$ %keeping alpha and eta constant, i.e.
keeping $\beta\hbar$ constant, which in these reduced units is equivalent to $\Phi\rightarrow\infty$ with constant $\alpha$ and $\eta_s$.
The classical rate for this system is found in the limit $\alpha\rightarrow0$ and is given by 
\begin{align}
\label{LZ}
k^{(s)}_{\textrm{cl}}=\frac{\beta \Delta^2}{\hbar} \sqrt{\frac{\pi}{\Phi}}\frac{1}{1+\eta_s}\, \eu{-\Phi}.
\end{align}
The same expression can be derived \cite{nonoscillatory,Rips1995ET} alternatively
from a Boltzmann average of
the Landau-Zener formula \cite{Landau1932LZ,Zener1932LZ} in the golden-rule limit.

\section{Specification of the GR-QTST Hamiltonian}
\label{sec:definitions}
The energy constraint employed by the GR-QTST method is defined by \cite{GRQTST} 
\begin{align}
	\sigma_\lambda(\mathbf{x}) &= \tfrac{2}{3}\beta(E_0^{\text{v}}-E_1^{\text{v}}),
\end{align}
where
$E_n^{\text{v}}$ is the virial energy estimator for an imaginary-time path on PES $n\in\{0,1\}$ and is defined as
\begin{subequations}
\begin{align}
E_{0}^{\text{v}} &= \frac{1}{2N_{0}}\sum_{i=1}^{N_{0}}\left[V_{0}^{(i)}+V_{0}^{(i-1)}+\thalf \mathsf{g}_0^{(i)}\cdot(\mathsf{x}^{(i)}-\mathsf{s})+\thalf \mathsf{g}_0^{(i-1)}\cdot(\mathsf{x}^{(i-1)}-\mathsf{s})\right]\\
E_{1}^{\text{v}} &= \frac{1}{2N_{1}}\sum_{i=N_0+1}^{N}\left[V_{1}^{(i)}+V_{1}^{(i-1)}+\thalf \mathsf{g}_1^{(i)}\cdot(\mathsf{x}^{(i)}-\mathsf{s})+\thalf \mathsf{g}_1^{(i-1)}\cdot(\mathsf{x}^{(i-1)}-\mathsf{s})\right] ,
\end{align}
\end{subequations}
in which $V_{n}^{(i)}\equiv V_{n}(\mathsf{x}^{(i)})$ and $\mathsf{g}_n^{(i)}\equiv\frac{\partial V_{n}^{(i)}}{\partial \mathsf{x}^{(i)}}$ is the gradient vector at bead $i$.
The reference coordinate is
$\mathsf{s} = \mathsf{x}_+ - \frac{\mathsf{g}^-}{\|\mathsf{g}^-\|^2}V^-$, where
$\mathsf{x}_+ \equiv \frac{\mathsf{x}^{(0)}+\mathsf{x}^{(N_0)}}{2}$ is the average position of the hopping beads, at which the potential difference is
$V^- \equiv V_0(\mathsf{x}_+)-V_1(\mathsf{x}_+)$ and $\mathsf{g}^- \equiv \frac{\partial V^-}{\partial \mathsf{x}_+}$.
All the vectors here are defined as column vectors.
Note that these functions are only defined for $0<\lambda<1$, i.e.\ not at the boundaries.

The first derivatives are
\begin{align}
\frac{\partial \sigma_\lambda}{\partial\textbf{x}}&=\frac{2}{3}\beta\left(\frac{\partial E_0^{\text{v}}}{\partial\textbf{x}}-\frac{\partial E_1^{\text{v}}}{\partial\textbf{x}}\right),
\\
\frac{\partial E^{\text{v}}_{0}}{\partial \mathsf{x}^{(i)}} &= 
	\begin{cases}
	\frac{1}{2N_{0}}\left[3\mathsf{g}_0^{(i)} + \mathsf{H}_0^{(i)}(\mathsf{x}^{(i)}-\mathsf{s}) \right],~~~i\in\{1,\dots,N_{0}-1\}
	\\
	\frac{1}{2N_{0}}\left[\tfrac{3}{2}\mathsf{g}_0^{(i)} + \tfrac{1}{2} \mathsf{H}_0^{(i)}(\mathsf{x}^{(i)}-\mathsf{s}) - \left.\frac{\partial\mathsf{s}}{\partial\mathsf{x}^{(i)}}\right. \cdot
	\left(\sum_{i'=1}^{N_{0}}\tfrac{1}{2}\mathsf{g}_0^{(i')} + \tfrac{1}{2}\mathsf{g}_0^{(i'-1)}\right) \right],~~~i\in\{0,N_{0}\}
	\\
	0,~~~i\in\{N_0+1,\dots,N-1\}
	\end{cases}
\\
\frac{\partial s_j}{\partial x^{(0)}_k} &= \frac{\partial s_j}{\partial x^{(N_{0})}_k} = \frac{1}{2}\left[\delta_{jk} - \frac{g^-_j g^-_k}{\|\mathsf{g}^-\|^2} - V^- \left(\frac{H^-_{jk}}{\|\mathsf{g}^-\|^2} - \frac{2g^-_j}{\|\mathsf{g}^-\|^4}(\mathsf{H}^-\mathsf{g}^-)_{k} \right) \right],
\label{dsdx}
\end{align}
where $\mathsf{H}_n^{(i)}=\frac{\partial^2 V_{n}^{(i)}}{\partial \mathsf{x}^{(i)}\partial\mathsf{x}^{(i)}}$ is the Hessian matrix at bead $i$ and $\mathsf{H}^-=\frac{\partial^2 V^-}{\partial \mathsf{x}_+\partial\mathsf{x}_+}$. %, and $\textbf{H}^-_v$ is the $v$th column of $\textbf{H}^-$.
In \eqn{dsdx}, $j$ is the row index and $k$ is the column index.
The derivatives of $E_1^\text{v}$ are defined in the same way as those of $E_0^\text{v}$.
Note that the explicit calculation of the Hessian matrix in these functions could be circumvented in the future with the
finite difference method described in \Ref{Kapil2016PI}.
\section{Replica-exchange}
\label{sec:replica}
Replica exchange was used in the PIMD simulations, with exchanges between ensembles $k$ and $k'$ accepted with the probability:
\begin{equation}
    A=\text{min}\{1,\eu{(\beta_N^{[k]}-\beta_N^{[k']})(U(\textbf{x}^{[k]})-U(\textbf{x}^{[k']}))}\},
\end{equation}
in which $U=U_{\text{RP}}+U_N^{(\lambda)}$.
Note that the function $U$ remains the same for all the replicas ($\omega_N$ does not change), simply corresponding to the replica exchange of a classical molecular dynamics simulation \cite{DaanFrenkel_book,B509983H} applied to the ring-polymer phase space.
One could alternatively do ``quantum" replica-exchange in PIMD allowing the spring term in $U_\text{RP}$ to change according to the replica temperature \cite{doi:10.1021/ct500447r}, but 
as we do not need quantum statistics at higher temperatures, this approach was unnecessary for this work, and was also not found to be any more efficient.
Swaps are attempted between systems with adjacent temperatures ($k' = k + 1$) every 25 steps.
The temperatures of $M$ replicas were chosen as $\beta_N^{[1]}=\beta_N$, where $\beta_N^{[M]}$ is the inverse temperature of the hottest replica (values given in \sref{results}), and $(\beta_N^{[k]})^{1/p}$ are distributed linearly between $(\beta_N^{[1]})^{1/p}$ and $(\beta_N^{[M]})^{1/p}$.
We gave $p$ a value of 4 or 5, which leads to a replica-exchange probability of 20--50\% in all the simulations in this work. 

\bibliography{main.bbl}

\begin{thebibliography}{89}%
\makeatletter
\providecommand \@ifxundefined [1]{%
 \@ifx{#1\undefined}
}%
\providecommand \@ifnum [1]{%
 \ifnum #1\expandafter \@firstoftwo
 \else \expandafter \@secondoftwo
 \fi
}%
\providecommand \@ifx [1]{%
 \ifx #1\expandafter \@firstoftwo
 \else \expandafter \@secondoftwo
 \fi
}%
\providecommand \natexlab [1]{#1}%
\providecommand \enquote  [1]{``#1''}%
\providecommand \bibnamefont  [1]{#1}%
\providecommand \bibfnamefont [1]{#1}%
\providecommand \citenamefont [1]{#1}%
\providecommand \href@noop [0]{\@secondoftwo}%
\providecommand \href [0]{\begingroup \@sanitize@url \@href}%
\providecommand \@href[1]{\@@startlink{#1}\@@href}%
\providecommand \@@href[1]{\endgroup#1\@@endlink}%
\providecommand \@sanitize@url [0]{\catcode `\\12\catcode `\$12\catcode
  `\&12\catcode `\#12\catcode `\^12\catcode `\_12\catcode `\%12\relax}%
\providecommand \@@startlink[1]{}%
\providecommand \@@endlink[0]{}%
\providecommand \url  [0]{\begingroup\@sanitize@url \@url }%
\providecommand \@url [1]{\endgroup\@href {#1}{\urlprefix }}%
\providecommand \urlprefix  [0]{URL }%
\providecommand \Eprint [0]{\href }%
\providecommand \doibase [0]{http://dx.doi.org/}%
\providecommand \selectlanguage [0]{\@gobble}%
\providecommand \bibinfo  [0]{\@secondoftwo}%
\providecommand \bibfield  [0]{\@secondoftwo}%
\providecommand \translation [1]{[#1]}%
\providecommand \BibitemOpen [0]{}%
\providecommand \bibitemStop [0]{}%
\providecommand \bibitemNoStop [0]{.\EOS\space}%
\providecommand \EOS [0]{\spacefactor3000\relax}%
\providecommand \BibitemShut  [1]{\csname bibitem#1\endcsname}%
\let\auto@bib@innerbib\@empty
%</preamble>
\bibitem [{\citenamefont {Blumberger}(2015)}]{Blumberger2015chemrev}%
  \BibitemOpen
  \bibfield  {author} {\bibinfo {author} {\bibfnamefont {J.}~\bibnamefont
  {Blumberger}},\ }\href {\doibase 10.1021/acs.chemrev.5b00298} {\bibfield
  {journal} {\bibinfo  {journal} {Chem. Rev.}\ }\textbf {\bibinfo {volume}
  {115}},\ \bibinfo {pages} {11191} (\bibinfo {year} {2015})}\BibitemShut
  {NoStop}%
\bibitem [{\citenamefont {Friis}\ \emph {et~al.}(1999)\citenamefont {Friis},
  \citenamefont {Andersen}, \citenamefont {Kharkats}, \citenamefont
  {Kuznetsov}, \citenamefont {Nichols}, \citenamefont {Zhang},\ and\
  \citenamefont {Ulstrup}}]{Friis1379}%
  \BibitemOpen
  \bibfield  {author} {\bibinfo {author} {\bibfnamefont {E.~P.}\ \bibnamefont
  {Friis}}, \bibinfo {author} {\bibfnamefont {J.~E.~T.}\ \bibnamefont
  {Andersen}}, \bibinfo {author} {\bibfnamefont {Y.~I.}\ \bibnamefont
  {Kharkats}}, \bibinfo {author} {\bibfnamefont {A.~M.}\ \bibnamefont
  {Kuznetsov}}, \bibinfo {author} {\bibfnamefont {R.~J.}\ \bibnamefont
  {Nichols}}, \bibinfo {author} {\bibfnamefont {J.-D.}\ \bibnamefont {Zhang}},
  \ and\ \bibinfo {author} {\bibfnamefont {J.}~\bibnamefont {Ulstrup}},\ }\href
  {\doibase 10.1073/pnas.96.4.1379} {\bibfield  {journal} {\bibinfo  {journal}
  {Proc. Natl. Acad. Sci. U.S.A.}\ }\textbf {\bibinfo {volume} {96}},\ \bibinfo
  {pages} {1379} (\bibinfo {year} {1999})},\ \Eprint
  {http://arxiv.org/abs/https://www.pnas.org/content/96/4/1379.full.pdf}
  {https://www.pnas.org/content/96/4/1379.full.pdf} \BibitemShut {NoStop}%
\bibitem [{\citenamefont {Pia}\ \emph {et~al.}(2011)\citenamefont {Pia},
  \citenamefont {Chi}, \citenamefont {Jones}, \citenamefont {Macdonald},
  \citenamefont {Ulstrup},\ and\ \citenamefont
  {Elliott}}]{doi:10.1021/nl103334q}%
  \BibitemOpen
  \bibfield  {author} {\bibinfo {author} {\bibfnamefont {E.~A.~D.}\
  \bibnamefont {Pia}}, \bibinfo {author} {\bibfnamefont {Q.}~\bibnamefont
  {Chi}}, \bibinfo {author} {\bibfnamefont {D.~D.}\ \bibnamefont {Jones}},
  \bibinfo {author} {\bibfnamefont {J.~E.}\ \bibnamefont {Macdonald}}, \bibinfo
  {author} {\bibfnamefont {J.}~\bibnamefont {Ulstrup}}, \ and\ \bibinfo
  {author} {\bibfnamefont {M.}~\bibnamefont {Elliott}},\ }\href {\doibase
  10.1021/nl103334q} {\bibfield  {journal} {\bibinfo  {journal} {Nano Lett.}\
  }\textbf {\bibinfo {volume} {11}},\ \bibinfo {pages} {176} (\bibinfo {year}
  {2011})},\ \Eprint {http://arxiv.org/abs/https://doi.org/10.1021/nl103334q}
  {https://doi.org/10.1021/nl103334q} \BibitemShut {NoStop}%
\bibitem [{\citenamefont {Art\'{e}s}\ \emph {et~al.}(2011)\citenamefont
  {Art\'{e}s}, \citenamefont {D\'{i}ez-P\'{e}rez}, \citenamefont {Sanz},\ and\
  \citenamefont {Gorostiza}}]{doi:10.1021/nn103236e}%
  \BibitemOpen
  \bibfield  {author} {\bibinfo {author} {\bibfnamefont {J.~M.}\ \bibnamefont
  {Art\'{e}s}}, \bibinfo {author} {\bibfnamefont {I.}~\bibnamefont
  {D\'{i}ez-P\'{e}rez}}, \bibinfo {author} {\bibfnamefont {F.}~\bibnamefont
  {Sanz}}, \ and\ \bibinfo {author} {\bibfnamefont {P.}~\bibnamefont
  {Gorostiza}},\ }\href {\doibase 10.1021/nn103236e} {\bibfield  {journal}
  {\bibinfo  {journal} {ACS Nano}\ }\textbf {\bibinfo {volume} {5}},\ \bibinfo
  {pages} {2060} (\bibinfo {year} {2011})},\ \Eprint
  {http://arxiv.org/abs/https://doi.org/10.1021/nn103236e}
  {https://doi.org/10.1021/nn103236e} \BibitemShut {NoStop}%
\bibitem [{\citenamefont {Chandler}(1998)}]{ChandlerET}%
  \BibitemOpen
  \bibfield  {author} {\bibinfo {author} {\bibfnamefont {D.}~\bibnamefont
  {Chandler}},\ }in\ \href@noop {} {\emph {\bibinfo {booktitle} {Classical and
  Quantum Dynamics in Condensed Phase Simulations}}},\ \bibinfo {editor}
  {edited by\ \bibinfo {editor} {\bibfnamefont {B.~J.}\ \bibnamefont {Berne}},
  \bibinfo {editor} {\bibfnamefont {G.}~\bibnamefont {Ciccotti}}, \ and\
  \bibinfo {editor} {\bibfnamefont {D.~F.}\ \bibnamefont {Coker}}}\ (\bibinfo
  {publisher} {World Scientific},\ \bibinfo {address} {Singapore},\ \bibinfo
  {year} {1998})\ Chap.~\bibinfo {chapter} {2}, pp.\ \bibinfo {pages}
  {25--49}\BibitemShut {NoStop}%
\bibitem [{\citenamefont {Fermi}(1950)}]{Fermi}%
  \BibitemOpen
  \bibfield  {author} {\bibinfo {author} {\bibfnamefont {E.}~\bibnamefont
  {Fermi}},\ }\href@noop {} {\emph {\bibinfo {title} {Nuclear Physics}}}\
  (\bibinfo  {publisher} {University of Chicago Press, Chicago, USA},\ \bibinfo
  {year} {1950})\BibitemShut {NoStop}%
\bibitem [{\citenamefont {Marcus}(1964)}]{Marcus1964review}%
  \BibitemOpen
  \bibfield  {author} {\bibinfo {author} {\bibfnamefont {R.~A.}\ \bibnamefont
  {Marcus}},\ }\href {\doibase 10.1146/annurev.pc.15.100164.001103} {\bibfield
  {journal} {\bibinfo  {journal} {Annu. Rev. Phys. Chem.}\ }\textbf {\bibinfo
  {volume} {15}},\ \bibinfo {pages} {155} (\bibinfo {year} {1964})}\BibitemShut
  {NoStop}%
\bibitem [{\citenamefont {Marcus}(1993)}]{Marcus1993review}%
  \BibitemOpen
  \bibfield  {author} {\bibinfo {author} {\bibfnamefont {R.~A.}\ \bibnamefont
  {Marcus}},\ }\href {\doibase 10.1103/RevModPhys.65.599} {\bibfield  {journal}
  {\bibinfo  {journal} {Rev. Mod. Phys.}\ }\textbf {\bibinfo {volume} {65}},\
  \bibinfo {pages} {599} (\bibinfo {year} {1993})}\BibitemShut {NoStop}%
\bibitem [{\citenamefont
  {Wolynes}(1987{\natexlab{a}})}]{Wolynes1987nonadiabatic}%
  \BibitemOpen
  \bibfield  {author} {\bibinfo {author} {\bibfnamefont {P.~G.}\ \bibnamefont
  {Wolynes}},\ }\href {\doibase 10.1063/1.453440} {\bibfield  {journal}
  {\bibinfo  {journal} {J.~Chem. Phys.}\ }\textbf {\bibinfo {volume} {87}},\
  \bibinfo {pages} {6559} (\bibinfo {year} {1987}{\natexlab{a}})}\BibitemShut
  {NoStop}%
\bibitem [{\citenamefont
  {Wolynes}(1987{\natexlab{b}})}]{Wolynes1987dissipation}%
  \BibitemOpen
  \bibfield  {author} {\bibinfo {author} {\bibfnamefont {P.~G.}\ \bibnamefont
  {Wolynes}},\ }\href {\doibase 10.1063/1.452146} {\bibfield  {journal}
  {\bibinfo  {journal} {J.~Chem. Phys.}\ }\textbf {\bibinfo {volume} {86}},\
  \bibinfo {pages} {1957} (\bibinfo {year} {1987}{\natexlab{b}})}\BibitemShut
  {NoStop}%
\bibitem [{\citenamefont {Cline~Jr}\ and\ \citenamefont
  {Wolynes}(1987)}]{Cline1987nonadiabatic}%
  \BibitemOpen
  \bibfield  {author} {\bibinfo {author} {\bibfnamefont {R.~E.}\ \bibnamefont
  {Cline~Jr}}\ and\ \bibinfo {author} {\bibfnamefont {P.~G.}\ \bibnamefont
  {Wolynes}},\ }\href {\doibase 10.1063/1.451942} {\bibfield  {journal}
  {\bibinfo  {journal} {J.~Chem. Phys.}\ }\textbf {\bibinfo {volume} {86}},\
  \bibinfo {pages} {3836} (\bibinfo {year} {1987})}\BibitemShut {NoStop}%
\bibitem [{\citenamefont {Onuchic}\ and\ \citenamefont
  {Wolynes}(1988)}]{Onuchic1988rate}%
  \BibitemOpen
  \bibfield  {author} {\bibinfo {author} {\bibfnamefont {J.~N.}\ \bibnamefont
  {Onuchic}}\ and\ \bibinfo {author} {\bibfnamefont {P.~G.}\ \bibnamefont
  {Wolynes}},\ }\href {\doibase 10.1021/j100334a007} {\bibfield  {journal}
  {\bibinfo  {journal} {J.~Phys. Chem.}\ }\textbf {\bibinfo {volume} {92}},\
  \bibinfo {pages} {6495} (\bibinfo {year} {1988})}\BibitemShut {NoStop}%
\bibitem [{\citenamefont {Lawrence}\ and\ \citenamefont
  {Manolopoulos}(2018)}]{Lawrence2018Wolynes}%
  \BibitemOpen
  \bibfield  {author} {\bibinfo {author} {\bibfnamefont {J.~E.}\ \bibnamefont
  {Lawrence}}\ and\ \bibinfo {author} {\bibfnamefont {D.~E.}\ \bibnamefont
  {Manolopoulos}},\ }\href {\doibase 10.1063/1.5002894} {\bibfield  {journal}
  {\bibinfo  {journal} {J. Chem. Phys.}\ }\textbf {\bibinfo {volume} {148}},\
  \bibinfo {pages} {102313} (\bibinfo {year} {2018})}\BibitemShut {NoStop}%
\bibitem [{\citenamefont {Zhu}\ and\ \citenamefont
  {Nakamura}(1994)}]{Zhu1994ZN}%
  \BibitemOpen
  \bibfield  {author} {\bibinfo {author} {\bibfnamefont {C.}~\bibnamefont
  {Zhu}}\ and\ \bibinfo {author} {\bibfnamefont {H.}~\bibnamefont {Nakamura}},\
  }\href {\doibase 10.1063/1.467877} {\bibfield  {journal} {\bibinfo  {journal}
  {J.~Chem. Phys.}\ }\textbf {\bibinfo {volume} {101}},\ \bibinfo {pages}
  {10630} (\bibinfo {year} {1994})}\BibitemShut {NoStop}%
\bibitem [{\citenamefont {Cao}, \citenamefont {Minichino},\ and\ \citenamefont
  {Voth}(1995)}]{Cao1995nonadiabatic}%
  \BibitemOpen
  \bibfield  {author} {\bibinfo {author} {\bibfnamefont {J.}~\bibnamefont
  {Cao}}, \bibinfo {author} {\bibfnamefont {C.}~\bibnamefont {Minichino}}, \
  and\ \bibinfo {author} {\bibfnamefont {G.~A.}\ \bibnamefont {Voth}},\ }\href
  {\doibase 10.1063/1.469762} {\bibfield  {journal} {\bibinfo  {journal}
  {J.~Chem. Phys.}\ }\textbf {\bibinfo {volume} {103}},\ \bibinfo {pages}
  {1391} (\bibinfo {year} {1995})}\BibitemShut {NoStop}%
\bibitem [{\citenamefont {Cao}\ and\ \citenamefont
  {Voth}(1997)}]{Cao1997nonadiabatic}%
  \BibitemOpen
  \bibfield  {author} {\bibinfo {author} {\bibfnamefont {J.}~\bibnamefont
  {Cao}}\ and\ \bibinfo {author} {\bibfnamefont {G.~A.}\ \bibnamefont {Voth}},\
  }\href {\doibase 10.1063/1.474123} {\bibfield  {journal} {\bibinfo  {journal}
  {J.~Chem. Phys.}\ }\textbf {\bibinfo {volume} {106}},\ \bibinfo {pages}
  {1769} (\bibinfo {year} {1997})}\BibitemShut {NoStop}%
\bibitem [{\citenamefont {Schwieters}\ and\ \citenamefont
  {Voth}(1998)}]{Schwieters1998diabatic}%
  \BibitemOpen
  \bibfield  {author} {\bibinfo {author} {\bibfnamefont {C.~D.}\ \bibnamefont
  {Schwieters}}\ and\ \bibinfo {author} {\bibfnamefont {G.~A.}\ \bibnamefont
  {Voth}},\ }\href {\doibase 10.1063/1.475467} {\bibfield  {journal} {\bibinfo
  {journal} {J.~Chem. Phys.}\ }\textbf {\bibinfo {volume} {108}},\ \bibinfo
  {pages} {1055} (\bibinfo {year} {1998})}\BibitemShut {NoStop}%
\bibitem [{\citenamefont {Schwieters}\ and\ \citenamefont
  {Voth}(1999)}]{Schwieters1999diabatic}%
  \BibitemOpen
  \bibfield  {author} {\bibinfo {author} {\bibfnamefont {C.~D.}\ \bibnamefont
  {Schwieters}}\ and\ \bibinfo {author} {\bibfnamefont {G.~A.}\ \bibnamefont
  {Voth}},\ }\href {\doibase 10.1063/1.479569} {\bibfield  {journal} {\bibinfo
  {journal} {J.~Chem. Phys.}\ }\textbf {\bibinfo {volume} {111}},\ \bibinfo
  {pages} {2869} (\bibinfo {year} {1999})}\BibitemShut {NoStop}%
\bibitem [{\citenamefont {Hammes-Schiffer}\ and\ \citenamefont
  {Stuchebrukhov}(2010)}]{HammesSchiffer2010PCET}%
  \BibitemOpen
  \bibfield  {author} {\bibinfo {author} {\bibfnamefont {S.}~\bibnamefont
  {Hammes-Schiffer}}\ and\ \bibinfo {author} {\bibfnamefont {A.~A.}\
  \bibnamefont {Stuchebrukhov}},\ }\href {\doibase 10.1021/cr1001436}
  {\bibfield  {journal} {\bibinfo  {journal} {Chem. Rev.}\ }\textbf {\bibinfo
  {volume} {110}},\ \bibinfo {pages} {6939} (\bibinfo {year}
  {2010})}\BibitemShut {NoStop}%
\bibitem [{\citenamefont {Hammes-Schiffer}(2015)}]{HammesSchiffer2015PCET}%
  \BibitemOpen
  \bibfield  {author} {\bibinfo {author} {\bibfnamefont {S.}~\bibnamefont
  {Hammes-Schiffer}},\ }\href {\doibase 10.1021/jacs.5b04087} {\bibfield
  {journal} {\bibinfo  {journal} {J. Am. Chem. Soc.}\ }\textbf {\bibinfo
  {volume} {137}},\ \bibinfo {pages} {8860} (\bibinfo {year}
  {2015})}\BibitemShut {NoStop}%
\bibitem [{\citenamefont {Kretchmer}\ and\ \citenamefont
  {Miller~III}(2013)}]{Kretchmer2013ET}%
  \BibitemOpen
  \bibfield  {author} {\bibinfo {author} {\bibfnamefont {J.~S.}\ \bibnamefont
  {Kretchmer}}\ and\ \bibinfo {author} {\bibfnamefont {T.~F.}\ \bibnamefont
  {Miller~III}},\ }\href {\doibase 10.1063/1.4797462} {\bibfield  {journal}
  {\bibinfo  {journal} {J.~Chem. Phys.}\ }\textbf {\bibinfo {volume} {138}},\
  \bibinfo {pages} {134109} (\bibinfo {year} {2013})}\BibitemShut {NoStop}%
\bibitem [{\citenamefont {Menzeleev}, \citenamefont {Bell},\ and\ \citenamefont
  {Miller~III}(2014)}]{Menzeleev2014kinetic}%
  \BibitemOpen
  \bibfield  {author} {\bibinfo {author} {\bibfnamefont {A.~R.}\ \bibnamefont
  {Menzeleev}}, \bibinfo {author} {\bibfnamefont {F.}~\bibnamefont {Bell}}, \
  and\ \bibinfo {author} {\bibfnamefont {T.~F.}\ \bibnamefont {Miller~III}},\
  }\href {\doibase 10.1063/1.4863919} {\bibfield  {journal} {\bibinfo
  {journal} {J.~Chem. Phys.}\ }\textbf {\bibinfo {volume} {140}},\ \bibinfo
  {pages} {064103} (\bibinfo {year} {2014})}\BibitemShut {NoStop}%
\bibitem [{\citenamefont {Kretchmer}\ and\ \citenamefont
  {Miller~III}(2016)}]{Kretchmer2016KCRPMD}%
  \BibitemOpen
  \bibfield  {author} {\bibinfo {author} {\bibfnamefont {J.~S.}\ \bibnamefont
  {Kretchmer}}\ and\ \bibinfo {author} {\bibfnamefont {T.~F.}\ \bibnamefont
  {Miller~III}},\ }\href {\doibase 10.1039/c6fd00143b} {\bibfield  {journal}
  {\bibinfo  {journal} {Faraday Discuss.}\ }\textbf {\bibinfo {volume} {195}},\
  \bibinfo {pages} {191} (\bibinfo {year} {2016})}\BibitemShut {NoStop}%
\bibitem [{\citenamefont {Richardson}, \citenamefont {Bauer},\ and\
  \citenamefont {Thoss}(2015)}]{GoldenGreens}%
  \BibitemOpen
  \bibfield  {author} {\bibinfo {author} {\bibfnamefont {J.~O.}\ \bibnamefont
  {Richardson}}, \bibinfo {author} {\bibfnamefont {R.}~\bibnamefont {Bauer}}, \
  and\ \bibinfo {author} {\bibfnamefont {M.}~\bibnamefont {Thoss}},\ }\href
  {\doibase 10.1063/1.4932361} {\bibfield  {journal} {\bibinfo  {journal}
  {J.~Chem. Phys.}\ }\textbf {\bibinfo {volume} {143}},\ \bibinfo {pages}
  {134115} (\bibinfo {year} {2015})},\ \Eprint
  {http://arxiv.org/abs/1508.04919} {arXiv:1508.04919 [physics.chem-ph]}
  \BibitemShut {NoStop}%
\bibitem [{\citenamefont {Richardson}(2015)}]{GoldenRPI}%
  \BibitemOpen
  \bibfield  {author} {\bibinfo {author} {\bibfnamefont {J.~O.}\ \bibnamefont
  {Richardson}},\ }\href {\doibase 10.1063/1.4932362} {\bibfield  {journal}
  {\bibinfo  {journal} {J.~Chem. Phys.}\ }\textbf {\bibinfo {volume} {143}},\
  \bibinfo {pages} {134116} (\bibinfo {year} {2015})},\ \Eprint
  {http://arxiv.org/abs/1508.05195} {arXiv:1508.05195 [physics.chem-ph]}
  \BibitemShut {NoStop}%
\bibitem [{\citenamefont {Mattiat}\ and\ \citenamefont
  {Richardson}(2018)}]{AsymSysBath}%
  \BibitemOpen
  \bibfield  {author} {\bibinfo {author} {\bibfnamefont {J.}~\bibnamefont
  {Mattiat}}\ and\ \bibinfo {author} {\bibfnamefont {J.~O.}\ \bibnamefont
  {Richardson}},\ }\href {\doibase 10.1063/1.5001116} {\bibfield  {journal}
  {\bibinfo  {journal} {J. Chem. Phys.}\ }\textbf {\bibinfo {volume} {148}},\
  \bibinfo {pages} {102311} (\bibinfo {year} {2018})},\ \Eprint
  {http://arxiv.org/abs/1708.06702} {arXiv:1708.06702 [physics.chem-ph]}
  \BibitemShut {NoStop}%
\bibitem [{\citenamefont {Duke}\ and\ \citenamefont
  {Ananth}(2016)}]{Duke2016Faraday}%
  \BibitemOpen
  \bibfield  {author} {\bibinfo {author} {\bibfnamefont {J.~R.}\ \bibnamefont
  {Duke}}\ and\ \bibinfo {author} {\bibfnamefont {N.}~\bibnamefont {Ananth}},\
  }\href {\doibase 10.1039/C6FD00123H} {\bibfield  {journal} {\bibinfo
  {journal} {Faraday Discuss.}\ }\textbf {\bibinfo {volume} {195}},\ \bibinfo
  {pages} {253} (\bibinfo {year} {2016})}\BibitemShut {NoStop}%
\bibitem [{\citenamefont {Pierre}\ \emph {et~al.}(2017)\citenamefont {Pierre},
  \citenamefont {Duke}, \citenamefont {Hele},\ and\ \citenamefont
  {Ananth}}]{Pierre2017MVRPMD}%
  \BibitemOpen
  \bibfield  {author} {\bibinfo {author} {\bibfnamefont {S.}~\bibnamefont
  {Pierre}}, \bibinfo {author} {\bibfnamefont {J.~R.}\ \bibnamefont {Duke}},
  \bibinfo {author} {\bibfnamefont {T.~J.}\ \bibnamefont {Hele}}, \ and\
  \bibinfo {author} {\bibfnamefont {N.}~\bibnamefont {Ananth}},\ }\href
  {\doibase 10.1063/1.4986517} {\bibfield  {journal} {\bibinfo  {journal}
  {J.~Chem. Phys.}\ }\textbf {\bibinfo {volume} {147}},\ \bibinfo {pages}
  {234103} (\bibinfo {year} {2017})}\BibitemShut {NoStop}%
\bibitem [{\citenamefont {Tao}, \citenamefont {Shushkov},\ and\ \citenamefont
  {Miller~III}(2019)}]{Tao2019RPSH}%
  \BibitemOpen
  \bibfield  {author} {\bibinfo {author} {\bibfnamefont {X.}~\bibnamefont
  {Tao}}, \bibinfo {author} {\bibfnamefont {P.}~\bibnamefont {Shushkov}}, \
  and\ \bibinfo {author} {\bibfnamefont {T.~F.}\ \bibnamefont {Miller~III}},\
  }\href {\doibase 10.1021/acs.jpca.9b00877} {\bibfield  {journal} {\bibinfo
  {journal} {J.~Phys. Chem.~A}\ }\textbf {\bibinfo {volume} {123}},\ \bibinfo
  {pages} {3013} (\bibinfo {year} {2019})}\BibitemShut {NoStop}%
\bibitem [{\citenamefont {Curchod}\ and\ \citenamefont
  {Tavernelli}(2013)}]{doi:10.1063/1.4803835}%
  \BibitemOpen
  \bibfield  {author} {\bibinfo {author} {\bibfnamefont {B.~F.~E.}\
  \bibnamefont {Curchod}}\ and\ \bibinfo {author} {\bibfnamefont
  {I.}~\bibnamefont {Tavernelli}},\ }\href {\doibase 10.1063/1.4803835}
  {\bibfield  {journal} {\bibinfo  {journal} {J. Chem. Phys.}\ }\textbf
  {\bibinfo {volume} {138}},\ \bibinfo {pages} {184112} (\bibinfo {year}
  {2013})}\BibitemShut {NoStop}%
\bibitem [{\citenamefont {Abedi}, \citenamefont {Maitra},\ and\ \citenamefont
  {Gross}(2010)}]{PhysRevLett.105.123002}%
  \BibitemOpen
  \bibfield  {author} {\bibinfo {author} {\bibfnamefont {A.}~\bibnamefont
  {Abedi}}, \bibinfo {author} {\bibfnamefont {N.~T.}\ \bibnamefont {Maitra}}, \
  and\ \bibinfo {author} {\bibfnamefont {E.~K.~U.}\ \bibnamefont {Gross}},\
  }\href {\doibase 10.1103/PhysRevLett.105.123002} {\bibfield  {journal}
  {\bibinfo  {journal} {Phys. Rev. Lett.}\ }\textbf {\bibinfo {volume} {105}},\
  \bibinfo {pages} {123002} (\bibinfo {year} {2010})}\BibitemShut {NoStop}%
\bibitem [{\citenamefont {Min}\ \emph {et~al.}(2017)\citenamefont {Min},
  \citenamefont {Agostini}, \citenamefont {Tavernelli},\ and\ \citenamefont
  {Gross}}]{doi:10.1021/acs.jpclett.7b01249}%
  \BibitemOpen
  \bibfield  {author} {\bibinfo {author} {\bibfnamefont {S.~K.}\ \bibnamefont
  {Min}}, \bibinfo {author} {\bibfnamefont {F.}~\bibnamefont {Agostini}},
  \bibinfo {author} {\bibfnamefont {I.}~\bibnamefont {Tavernelli}}, \ and\
  \bibinfo {author} {\bibfnamefont {E.~K.~U.}\ \bibnamefont {Gross}},\ }\href
  {\doibase 10.1021/acs.jpclett.7b01249} {\bibfield  {journal} {\bibinfo
  {journal} {J. Chem. Phys. Lett.}\ }\textbf {\bibinfo {volume} {8}},\ \bibinfo
  {pages} {3048} (\bibinfo {year} {2017})}\BibitemShut {NoStop}%
\bibitem [{\citenamefont {Lu}\ and\ \citenamefont
  {Zhou}(2017)}]{doi:10.1063/1.4981021}%
  \BibitemOpen
  \bibfield  {author} {\bibinfo {author} {\bibfnamefont {J.}~\bibnamefont
  {Lu}}\ and\ \bibinfo {author} {\bibfnamefont {Z.}~\bibnamefont {Zhou}},\
  }\href {\doibase 10.1063/1.4981021} {\bibfield  {journal} {\bibinfo
  {journal} {J. Chem. Phys.}\ }\textbf {\bibinfo {volume} {146}},\ \bibinfo
  {pages} {154110} (\bibinfo {year} {2017})}\BibitemShut {NoStop}%
\bibitem [{\citenamefont {Shakib}\ and\ \citenamefont
  {Huo}(2017)}]{doi:10.1021/acs.jpclett.7b01343}%
  \BibitemOpen
  \bibfield  {author} {\bibinfo {author} {\bibfnamefont {F.~A.}\ \bibnamefont
  {Shakib}}\ and\ \bibinfo {author} {\bibfnamefont {P.}~\bibnamefont {Huo}},\
  }\href {\doibase 10.1021/acs.jpclett.7b01343} {\bibfield  {journal} {\bibinfo
   {journal} {J. Chem. Phys. Lett.}\ }\textbf {\bibinfo {volume} {8}},\
  \bibinfo {pages} {3073} (\bibinfo {year} {2017})}\BibitemShut {NoStop}%
\bibitem [{\citenamefont {Bader}, \citenamefont {Kuharski},\ and\ \citenamefont
  {Chandler}(1990)}]{Bader1990golden}%
  \BibitemOpen
  \bibfield  {author} {\bibinfo {author} {\bibfnamefont {J.~S.}\ \bibnamefont
  {Bader}}, \bibinfo {author} {\bibfnamefont {R.~A.}\ \bibnamefont {Kuharski}},
  \ and\ \bibinfo {author} {\bibfnamefont {D.}~\bibnamefont {Chandler}},\
  }\href {\doibase 10.1063/1.459596} {\bibfield  {journal} {\bibinfo  {journal}
  {J.~Chem. Phys.}\ }\textbf {\bibinfo {volume} {93}},\ \bibinfo {pages} {230}
  (\bibinfo {year} {1990})}\BibitemShut {NoStop}%
\bibitem [{\citenamefont {Siders}\ and\ \citenamefont
  {Marcus}(1981)}]{Siders1981inverted}%
  \BibitemOpen
  \bibfield  {author} {\bibinfo {author} {\bibfnamefont {P.}~\bibnamefont
  {Siders}}\ and\ \bibinfo {author} {\bibfnamefont {R.~A.}\ \bibnamefont
  {Marcus}},\ }\href {\doibase 10.1021/ja00394a004} {\bibfield  {journal}
  {\bibinfo  {journal} {J.~Am. Chem. Soc.}\ }\textbf {\bibinfo {volume}
  {103}},\ \bibinfo {pages} {748} (\bibinfo {year} {1981})}\BibitemShut
  {NoStop}%
\bibitem [{\citenamefont {Hammes-Schiffer}\ and\ \citenamefont
  {Soudackov}(2008)}]{HammesSchiffer2008PCET}%
  \BibitemOpen
  \bibfield  {author} {\bibinfo {author} {\bibfnamefont {S.}~\bibnamefont
  {Hammes-Schiffer}}\ and\ \bibinfo {author} {\bibfnamefont {A.~V.}\
  \bibnamefont {Soudackov}},\ }\href {\doibase 10.1021/jp805876e} {\bibfield
  {journal} {\bibinfo  {journal} {J.~Phys. Chem.~B}\ }\textbf {\bibinfo
  {volume} {112}},\ \bibinfo {pages} {14108} (\bibinfo {year}
  {2008})}\BibitemShut {NoStop}%
\bibitem [{\citenamefont {Hu}\ \emph {et~al.}(2014)\citenamefont {Hu},
  \citenamefont {Sharma}, \citenamefont {Scouras}, \citenamefont {Soudackov},
  \citenamefont {Carr}, \citenamefont {Hammes-Schiffer}, \citenamefont
  {Alber},\ and\ \citenamefont {Klinman}}]{Hu2014enzyme}%
  \BibitemOpen
  \bibfield  {author} {\bibinfo {author} {\bibfnamefont {S.}~\bibnamefont
  {Hu}}, \bibinfo {author} {\bibfnamefont {S.~C.}\ \bibnamefont {Sharma}},
  \bibinfo {author} {\bibfnamefont {A.~D.}\ \bibnamefont {Scouras}}, \bibinfo
  {author} {\bibfnamefont {A.~V.}\ \bibnamefont {Soudackov}}, \bibinfo {author}
  {\bibfnamefont {C.~A.~M.}\ \bibnamefont {Carr}}, \bibinfo {author}
  {\bibfnamefont {S.}~\bibnamefont {Hammes-Schiffer}}, \bibinfo {author}
  {\bibfnamefont {T.}~\bibnamefont {Alber}}, \ and\ \bibinfo {author}
  {\bibfnamefont {J.~P.}\ \bibnamefont {Klinman}},\ }\href {\doibase
  10.1021/ja502726s} {\bibfield  {journal} {\bibinfo  {journal} {J. Am. Chem.
  Soc.}\ }\textbf {\bibinfo {volume} {136}},\ \bibinfo {pages} {8157} (\bibinfo
  {year} {2014})}\BibitemShut {NoStop}%
\bibitem [{\citenamefont {Pollak}(2012)}]{Pollak2012tunnel}%
  \BibitemOpen
  \bibfield  {author} {\bibinfo {author} {\bibfnamefont {E.}~\bibnamefont
  {Pollak}},\ }\href {\doibase 10.1021/jp307556j} {\bibfield  {journal}
  {\bibinfo  {journal} {J.~Phys. Chem.~B}\ }\textbf {\bibinfo {volume} {116}},\
  \bibinfo {pages} {12966} (\bibinfo {year} {2012})}\BibitemShut {NoStop}%
\bibitem [{\citenamefont {Feynman}\ and\ \citenamefont
  {Hibbs}(1965)}]{Feynman}%
  \BibitemOpen
  \bibfield  {author} {\bibinfo {author} {\bibfnamefont {R.~P.}\ \bibnamefont
  {Feynman}}\ and\ \bibinfo {author} {\bibfnamefont {A.~R.}\ \bibnamefont
  {Hibbs}},\ }\href@noop {} {\emph {\bibinfo {title} {Quantum Mechanics and
  Path Integrals}}}\ (\bibinfo  {publisher} {McGraw-Hill},\ \bibinfo {address}
  {New York},\ \bibinfo {year} {1965})\BibitemShut {NoStop}%
\bibitem [{\citenamefont {Miller}(1975)}]{Miller1975semiclassical}%
  \BibitemOpen
  \bibfield  {author} {\bibinfo {author} {\bibfnamefont {W.~H.}\ \bibnamefont
  {Miller}},\ }\href {\doibase 10.1063/1.430676} {\bibfield  {journal}
  {\bibinfo  {journal} {J.~Chem. Phys.}\ }\textbf {\bibinfo {volume} {62}},\
  \bibinfo {pages} {1899} (\bibinfo {year} {1975})}\BibitemShut {NoStop}%
\bibitem [{\citenamefont {Coleman}(1977)}]{Uses_of_Instantons}%
  \BibitemOpen
  \bibfield  {author} {\bibinfo {author} {\bibfnamefont {S.}~\bibnamefont
  {Coleman}},\ }in\ \href@noop {} {\emph {\bibinfo {booktitle} {Proc. Int.
  School of Subnuclear Physics}}}\ (\bibinfo {organization} {Erice},\ \bibinfo
  {year} {1977})\ \bibinfo {note} {also in S. Coleman, \emph{Aspects of
  Symmetry}, chapter 7, pp. 265--350 (Cambridge University Press,
  1985)}\BibitemShut {NoStop}%
\bibitem [{\citenamefont {Althorpe}(2011)}]{Althorpe2011ImF}%
  \BibitemOpen
  \bibfield  {author} {\bibinfo {author} {\bibfnamefont {S.~C.}\ \bibnamefont
  {Althorpe}},\ }\href {\doibase 10.1063/1.3563045} {\bibfield  {journal}
  {\bibinfo  {journal} {J.~Chem. Phys.}\ }\textbf {\bibinfo {volume} {134}},\
  \bibinfo {pages} {114104} (\bibinfo {year} {2011})}\BibitemShut {NoStop}%
\bibitem [{\citenamefont {Richardson}(2016{\natexlab{a}})}]{AdiabaticGreens}%
  \BibitemOpen
  \bibfield  {author} {\bibinfo {author} {\bibfnamefont {J.~O.}\ \bibnamefont
  {Richardson}},\ }\href {\doibase 10.1063/1.4943866} {\bibfield  {journal}
  {\bibinfo  {journal} {J.~Chem. Phys.}\ }\textbf {\bibinfo {volume} {144}},\
  \bibinfo {pages} {114106} (\bibinfo {year} {2016}{\natexlab{a}})},\ \Eprint
  {http://arxiv.org/abs/1512.04292} {arXiv:1512.04292 [physics.chem-ph]}
  \BibitemShut {NoStop}%
\bibitem [{\citenamefont {Richardson}(2018{\natexlab{a}})}]{InstReview}%
  \BibitemOpen
  \bibfield  {author} {\bibinfo {author} {\bibfnamefont {J.~O.}\ \bibnamefont
  {Richardson}},\ }\href {\doibase 10.1080/0144235X.2018.1472353} {\bibfield
  {journal} {\bibinfo  {journal} {Int. Rev. Phys. Chem.}\ }\textbf {\bibinfo
  {volume} {37}},\ \bibinfo {pages} {171} (\bibinfo {year}
  {2018}{\natexlab{a}})}\BibitemShut {NoStop}%
\bibitem [{\citenamefont {Andersson}\ \emph {et~al.}(2009)\citenamefont
  {Andersson}, \citenamefont {Nyman}, \citenamefont {Arnaldsson}, \citenamefont
  {Manthe},\ and\ \citenamefont {J{\'o}nsson}}]{Andersson2009Hmethane}%
  \BibitemOpen
  \bibfield  {author} {\bibinfo {author} {\bibfnamefont {S.}~\bibnamefont
  {Andersson}}, \bibinfo {author} {\bibfnamefont {G.}~\bibnamefont {Nyman}},
  \bibinfo {author} {\bibfnamefont {A.}~\bibnamefont {Arnaldsson}}, \bibinfo
  {author} {\bibfnamefont {U.}~\bibnamefont {Manthe}}, \ and\ \bibinfo {author}
  {\bibfnamefont {H.}~\bibnamefont {J{\'o}nsson}},\ }\href {\doibase
  10.1021/jp811070w} {\bibfield  {journal} {\bibinfo  {journal} {J.~Phys.
  Chem.~A}\ }\textbf {\bibinfo {volume} {113}},\ \bibinfo {pages} {4468}
  (\bibinfo {year} {2009})}\BibitemShut {NoStop}%
\bibitem [{\citenamefont {Richardson}\ and\ \citenamefont
  {Althorpe}(2009)}]{RPInst}%
  \BibitemOpen
  \bibfield  {author} {\bibinfo {author} {\bibfnamefont {J.~O.}\ \bibnamefont
  {Richardson}}\ and\ \bibinfo {author} {\bibfnamefont {S.~C.}\ \bibnamefont
  {Althorpe}},\ }\href {\doibase 10.1063/1.3267318} {\bibfield  {journal}
  {\bibinfo  {journal} {J.~Chem. Phys.}\ }\textbf {\bibinfo {volume} {131}},\
  \bibinfo {pages} {214106} (\bibinfo {year} {2009})}\BibitemShut {NoStop}%
\bibitem [{\citenamefont {Richardson}(2018{\natexlab{b}})}]{Perspective}%
  \BibitemOpen
  \bibfield  {author} {\bibinfo {author} {\bibfnamefont {J.~O.}\ \bibnamefont
  {Richardson}},\ }\href {\doibase 10.1063/1.5028352} {\bibfield  {journal}
  {\bibinfo  {journal} {J. Chem. Phys.}\ }\textbf {\bibinfo {volume} {148}},\
  \bibinfo {pages} {200901} (\bibinfo {year} {2018}{\natexlab{b}})}\BibitemShut
  {NoStop}%
\bibitem [{\citenamefont {Chandler}\ and\ \citenamefont
  {Wolynes}(1981)}]{Chandler+Wolynes1981}%
  \BibitemOpen
  \bibfield  {author} {\bibinfo {author} {\bibfnamefont {D.}~\bibnamefont
  {Chandler}}\ and\ \bibinfo {author} {\bibfnamefont {P.~G.}\ \bibnamefont
  {Wolynes}},\ }\href {\doibase 10.1063/1.441588} {\bibfield  {journal}
  {\bibinfo  {journal} {J.~Chem. Phys.}\ }\textbf {\bibinfo {volume} {74}},\
  \bibinfo {pages} {4078} (\bibinfo {year} {1981})}\BibitemShut {NoStop}%
\bibitem [{\citenamefont {Parrinello}\ and\ \citenamefont
  {Rahman}(1984)}]{Parrinello1984Fcenter}%
  \BibitemOpen
  \bibfield  {author} {\bibinfo {author} {\bibfnamefont {M.}~\bibnamefont
  {Parrinello}}\ and\ \bibinfo {author} {\bibfnamefont {A.}~\bibnamefont
  {Rahman}},\ }\href {\doibase 10.1063/1.446740} {\bibfield  {journal}
  {\bibinfo  {journal} {J.~Chem. Phys.}\ }\textbf {\bibinfo {volume} {80}},\
  \bibinfo {pages} {860} (\bibinfo {year} {1984})}\BibitemShut {NoStop}%
\bibitem [{\citenamefont {Markland}\ and\ \citenamefont
  {Ceriotti}(2018)}]{Markland2018review}%
  \BibitemOpen
  \bibfield  {author} {\bibinfo {author} {\bibfnamefont {T.~E.}\ \bibnamefont
  {Markland}}\ and\ \bibinfo {author} {\bibfnamefont {M.}~\bibnamefont
  {Ceriotti}},\ }\href {\doibase 10.1038/s41570-017-0109} {\bibfield  {journal}
  {\bibinfo  {journal} {Nat. Rev. Chem.}\ }\textbf {\bibinfo {volume} {2}},\
  \bibinfo {pages} {0109} (\bibinfo {year} {2018})}\BibitemShut {NoStop}%
\bibitem [{\citenamefont {Zheng}, \citenamefont {McCammon},\ and\ \citenamefont
  {Wolynes}(1989)}]{Zheng1989ET}%
  \BibitemOpen
  \bibfield  {author} {\bibinfo {author} {\bibfnamefont {C.}~\bibnamefont
  {Zheng}}, \bibinfo {author} {\bibfnamefont {J.~A.}\ \bibnamefont {McCammon}},
  \ and\ \bibinfo {author} {\bibfnamefont {P.~G.}\ \bibnamefont {Wolynes}},\
  }\href {\doibase 10.1073/pnas.86.17.6441} {\bibfield  {journal} {\bibinfo
  {journal} {P. Natl. Acad. Sci. USA}\ }\textbf {\bibinfo {volume} {86}},\
  \bibinfo {pages} {6441} (\bibinfo {year} {1989})}\BibitemShut {NoStop}%
\bibitem [{\citenamefont {Zheng}, \citenamefont {McCammon},\ and\ \citenamefont
  {Wolynes}(1991)}]{Zheng1991ET}%
  \BibitemOpen
  \bibfield  {author} {\bibinfo {author} {\bibfnamefont {C.}~\bibnamefont
  {Zheng}}, \bibinfo {author} {\bibfnamefont {J.~A.}\ \bibnamefont {McCammon}},
  \ and\ \bibinfo {author} {\bibfnamefont {P.~G.}\ \bibnamefont {Wolynes}},\
  }\href {\doibase 10.1016/0301-0104(91)87070-C} {\bibfield  {journal}
  {\bibinfo  {journal} {Chem. Phys.}\ }\textbf {\bibinfo {volume} {158}},\
  \bibinfo {pages} {261} (\bibinfo {year} {1991})}\BibitemShut {NoStop}%
\bibitem [{\citenamefont {Thapa}, \citenamefont {Fang},\ and\ \citenamefont
  {Richardson}(2019)}]{GRQTST}%
  \BibitemOpen
  \bibfield  {author} {\bibinfo {author} {\bibfnamefont {M.~J.}\ \bibnamefont
  {Thapa}}, \bibinfo {author} {\bibfnamefont {W.}~\bibnamefont {Fang}}, \ and\
  \bibinfo {author} {\bibfnamefont {J.~O.}\ \bibnamefont {Richardson}},\ }\href
  {\doibase 10.1063/1.5081108} {\bibfield  {journal} {\bibinfo  {journal} {J.
  Chem. Phys.}\ }\textbf {\bibinfo {volume} {150}},\ \bibinfo {pages} {104107}
  (\bibinfo {year} {2019})},\ \Eprint {http://arxiv.org/abs/1811.05874}
  {arXiv:1811.05874 [physics.chem-ph]} \BibitemShut {NoStop}%
\bibitem [{\citenamefont {Hele}\ and\ \citenamefont
  {Althorpe}(2013)}]{Hele2013QTST}%
  \BibitemOpen
  \bibfield  {author} {\bibinfo {author} {\bibfnamefont {T.~J.~H.}\
  \bibnamefont {Hele}}\ and\ \bibinfo {author} {\bibfnamefont {S.~C.}\
  \bibnamefont {Althorpe}},\ }\href {\doibase 10.1063/1.4792697} {\bibfield
  {journal} {\bibinfo  {journal} {J.~Chem. Phys.}\ }\textbf {\bibinfo {volume}
  {138}},\ \bibinfo {pages} {084108} (\bibinfo {year} {2013})}\BibitemShut
  {NoStop}%
\bibitem [{\citenamefont {Mills}\ \emph {et~al.}(1997)\citenamefont {Mills},
  \citenamefont {Schenter}, \citenamefont {Makarov},\ and\ \citenamefont
  {J{\'o}nsson}}]{Mills1997QTST}%
  \BibitemOpen
  \bibfield  {author} {\bibinfo {author} {\bibfnamefont {G.}~\bibnamefont
  {Mills}}, \bibinfo {author} {\bibfnamefont {G.~K.}\ \bibnamefont {Schenter}},
  \bibinfo {author} {\bibfnamefont {D.~E.}\ \bibnamefont {Makarov}}, \ and\
  \bibinfo {author} {\bibfnamefont {H.}~\bibnamefont {J{\'o}nsson}},\ }\href
  {\doibase 10.1016/S0009-2614(97)00886-5} {\bibfield  {journal} {\bibinfo
  {journal} {Chem. Phys. Lett.}\ }\textbf {\bibinfo {volume} {278}},\ \bibinfo
  {pages} {91} (\bibinfo {year} {1997})}\BibitemShut {NoStop}%
\bibitem [{\citenamefont {Vaillant}\ \emph {et~al.}(2019)\citenamefont
  {Vaillant}, \citenamefont {Van\'{\i}{\v{c}}ek}, \citenamefont {Thapa},\ and\
  \citenamefont {Richardson}}]{QInst}%
  \BibitemOpen
  \bibfield  {author} {\bibinfo {author} {\bibfnamefont {C.~L.}\ \bibnamefont
  {Vaillant}}, \bibinfo {author} {\bibfnamefont {J.}~\bibnamefont
  {Van\'{\i}{\v{c}}ek}}, \bibinfo {author} {\bibfnamefont {M.~J.}\ \bibnamefont
  {Thapa}}, \ and\ \bibinfo {author} {\bibfnamefont {J.~O.}\ \bibnamefont
  {Richardson}},\ }\href@noop {} {\bibfield  {journal} {\bibinfo  {journal} {J.
  Chem. Phys.}\ } (\bibinfo {year} {2019})},\ \bibinfo {note} {in press},\
  \Eprint {http://arxiv.org/abs/1908.03419} {arXiv:1908.03419
  [physics.chem-ph]} \BibitemShut {NoStop}%
\bibitem [{\citenamefont {Richardson}\ and\ \citenamefont
  {Thoss}(2014)}]{nonoscillatory}%
  \BibitemOpen
  \bibfield  {author} {\bibinfo {author} {\bibfnamefont {J.~O.}\ \bibnamefont
  {Richardson}}\ and\ \bibinfo {author} {\bibfnamefont {M.}~\bibnamefont
  {Thoss}},\ }\href {\doibase 10.1063/1.4892865} {\bibfield  {journal}
  {\bibinfo  {journal} {J.~Chem. Phys.}\ }\textbf {\bibinfo {volume} {141}},\
  \bibinfo {pages} {074106} (\bibinfo {year} {2014})},\ \Eprint
  {http://arxiv.org/abs/1406.3144} {arXiv:1406.3144 [physics.chem-ph]}
  \BibitemShut {NoStop}%
\bibitem [{\citenamefont {Kretchmer}\ \emph {et~al.}(2018)\citenamefont
  {Kretchmer}, \citenamefont {Boekelheide}, \citenamefont {Warren},
  \citenamefont {Winkler}, \citenamefont {Gray},\ and\ \citenamefont
  {Miller}}]{Kretchmer2018KCRPMD}%
  \BibitemOpen
  \bibfield  {author} {\bibinfo {author} {\bibfnamefont {J.~S.}\ \bibnamefont
  {Kretchmer}}, \bibinfo {author} {\bibfnamefont {N.}~\bibnamefont
  {Boekelheide}}, \bibinfo {author} {\bibfnamefont {J.~J.}\ \bibnamefont
  {Warren}}, \bibinfo {author} {\bibfnamefont {J.~R.}\ \bibnamefont {Winkler}},
  \bibinfo {author} {\bibfnamefont {H.~B.}\ \bibnamefont {Gray}}, \ and\
  \bibinfo {author} {\bibfnamefont {T.~F.}\ \bibnamefont {Miller}},\
  }\href@noop {} {\bibfield  {journal} {\bibinfo  {journal} {P. Natl. Acad.
  Sci. USA}\ }\textbf {\bibinfo {volume} {115}},\ \bibinfo {pages} {6129}
  (\bibinfo {year} {2018})}\BibitemShut {NoStop}%
\bibitem [{\citenamefont {Yamamoto}(1960)}]{Yamamoto1960rate}%
  \BibitemOpen
  \bibfield  {author} {\bibinfo {author} {\bibfnamefont {T.}~\bibnamefont
  {Yamamoto}},\ }\href {\doibase 10.1063/1.1731099} {\bibfield  {journal}
  {\bibinfo  {journal} {J.~Chem. Phys.}\ }\textbf {\bibinfo {volume} {33}},\
  \bibinfo {pages} {281} (\bibinfo {year} {1960})}\BibitemShut {NoStop}%
\bibitem [{\citenamefont {Miller}(1974)}]{Miller1974QTST}%
  \BibitemOpen
  \bibfield  {author} {\bibinfo {author} {\bibfnamefont {W.~H.}\ \bibnamefont
  {Miller}},\ }\href {\doibase 10.1063/1.1682181} {\bibfield  {journal}
  {\bibinfo  {journal} {J.~Chem. Phys.}\ }\textbf {\bibinfo {volume} {61}},\
  \bibinfo {pages} {1823} (\bibinfo {year} {1974})}\BibitemShut {NoStop}%
\bibitem [{\citenamefont {Miller}, \citenamefont {Schwartz},\ and\
  \citenamefont {Tromp}(1983)}]{Miller1983rate}%
  \BibitemOpen
  \bibfield  {author} {\bibinfo {author} {\bibfnamefont {W.~H.}\ \bibnamefont
  {Miller}}, \bibinfo {author} {\bibfnamefont {S.~D.}\ \bibnamefont
  {Schwartz}}, \ and\ \bibinfo {author} {\bibfnamefont {J.~W.}\ \bibnamefont
  {Tromp}},\ }\href {\doibase 10.1063/1.445581} {\bibfield  {journal} {\bibinfo
   {journal} {J.~Chem. Phys.}\ }\textbf {\bibinfo {volume} {79}},\ \bibinfo
  {pages} {4889} (\bibinfo {year} {1983})}\BibitemShut {NoStop}%
\bibitem [{Note1()}]{Note1}%
  \BibitemOpen
  \bibinfo {note} {The simplest way to go beyond this limit would be to use the
  approach described in Ref.~\protect \rev@citealpnum
  {Lawrence2019ET}}\BibitemShut {NoStop}%
\bibitem [{\citenamefont {Lawrence}\ \emph {et~al.}(2019)\citenamefont
  {Lawrence}, \citenamefont {Fletcher}, \citenamefont {Lindoy},\ and\
  \citenamefont {Manolopoulos}}]{Lawrence2019ET}%
  \BibitemOpen
  \bibfield  {author} {\bibinfo {author} {\bibfnamefont {J.~E.}\ \bibnamefont
  {Lawrence}}, \bibinfo {author} {\bibfnamefont {T.}~\bibnamefont {Fletcher}},
  \bibinfo {author} {\bibfnamefont {L.~P.}\ \bibnamefont {Lindoy}}, \ and\
  \bibinfo {author} {\bibfnamefont {D.~E.}\ \bibnamefont {Manolopoulos}},\
  }\href {\doibase 10.1063/1.5116800} {\bibfield  {journal} {\bibinfo
  {journal} {J.~Chem. Phys.}\ }\textbf {\bibinfo {volume} {151}},\ \bibinfo
  {pages} {114119} (\bibinfo {year} {2019})}\BibitemShut {NoStop}%
\bibitem [{\citenamefont {Weiss}(2012)}]{Weiss}%
  \BibitemOpen
  \bibfield  {author} {\bibinfo {author} {\bibfnamefont {U.}~\bibnamefont
  {Weiss}},\ }\href@noop {} {\emph {\bibinfo {title} {Quantum Dissipative
  Systems}}},\ \bibinfo {edition} {4th}\ ed.\ (\bibinfo  {publisher} {World
  Scientific},\ \bibinfo {address} {Singapore},\ \bibinfo {year}
  {2012})\BibitemShut {NoStop}%
\bibitem [{\citenamefont {Bender}\ and\ \citenamefont
  {Orszag}(1978)}]{BenderBook}%
  \BibitemOpen
  \bibfield  {author} {\bibinfo {author} {\bibfnamefont {C.~M.}\ \bibnamefont
  {Bender}}\ and\ \bibinfo {author} {\bibfnamefont {S.~A.}\ \bibnamefont
  {Orszag}},\ }\href@noop {} {\emph {\bibinfo {title} {Advanced Mathematical
  Methods for Scientists and Engineers}}}\ (\bibinfo  {publisher}
  {McGraw-Hill},\ \bibinfo {address} {New York},\ \bibinfo {year}
  {1978})\BibitemShut {NoStop}%
\bibitem [{\citenamefont {Gutzwiller}(1990)}]{GutzwillerBook}%
  \BibitemOpen
  \bibfield  {author} {\bibinfo {author} {\bibfnamefont {M.~C.}\ \bibnamefont
  {Gutzwiller}},\ }\href {\doibase 10.1007/978-1-4612-0983-6} {\emph {\bibinfo
  {title} {Chaos in Classical and Quantum Mechanics}}}\ (\bibinfo  {publisher}
  {Springer-Verlag},\ \bibinfo {address} {New York},\ \bibinfo {year}
  {1990})\BibitemShut {NoStop}%
\bibitem [{Note2()}]{Note2}%
  \BibitemOpen
  \bibinfo {note} {Typically $\tau ^*$ is in the range $[0,\beta \hbar ]$, but
  $\phi _\protect \text {u}(\tau )$ can be analytically continued beyond this
  range to treat the inverted regime as described in Ref.~\protect
  \rev@citealpnum {Lawrence2018Wolynes}}\BibitemShut {NoStop}%
\bibitem [{\citenamefont {Richardson}(2016{\natexlab{b}})}]{Faraday}%
  \BibitemOpen
  \bibfield  {author} {\bibinfo {author} {\bibfnamefont {J.~O.}\ \bibnamefont
  {Richardson}},\ }\href {\doibase 10.1039/C6FD00119J} {\bibfield  {journal}
  {\bibinfo  {journal} {Faraday Discuss.}\ }\textbf {\bibinfo {volume} {195}},\
  \bibinfo {pages} {49} (\bibinfo {year} {2016}{\natexlab{b}})}\BibitemShut
  {NoStop}%
\bibitem [{Note3()}]{Note3}%
  \BibitemOpen
  \bibinfo {note} {This can be proved by setting the derivative of Eq.\protect
  \tmspace +\thinmuskip {.1667em}\protect \textup {\hbox {\mathsurround \z@
  \protect \normalfont (\ignorespaces \ref {distribution}\unskip \@@italiccorr
  )}} to 0, which requires the derivatives of $\phi _\protect \text
  {u}^{(\protect \text {A})}(\tau )$ and $\phi _\protect \text {u}^{(\protect
  \text {B})}(\tau )$ to have opposite signs.}\BibitemShut {Stop}%
\bibitem [{Note4()}]{Note4}%
  \BibitemOpen
  \bibinfo {note} {Further changes to the GR-QTST ansatz, such as always
  choosing $\tau =\beta \hbar /2$ or averaging over a fixed range of values,
  would in fact make it rigorously separable in this case.}\BibitemShut {Stop}%
\bibitem [{Note5()}]{Note5}%
  \BibitemOpen
  \bibinfo {note} {We employed the standard normal-mode update scheme which we
  found to be stable. In case of instability problems in future applications
  one could instead apply the scheme described in Ref.~\protect \rev@citealpnum
  {Korol2019Cayley}.}\BibitemShut {Stop}%
\bibitem [{\citenamefont {Korol}, \citenamefont {Bou-Rabee},\ and\
  \citenamefont {Miller~III}(2019)}]{Korol2019Cayley}%
  \BibitemOpen
  \bibfield  {author} {\bibinfo {author} {\bibfnamefont {R.}~\bibnamefont
  {Korol}}, \bibinfo {author} {\bibfnamefont {N.}~\bibnamefont {Bou-Rabee}}, \
  and\ \bibinfo {author} {\bibfnamefont {T.~F.}\ \bibnamefont {Miller~III}},\
  }\href@noop {} {\bibfield  {journal} {\bibinfo  {journal} {J.~Chem. Phys.}\
  }\textbf {\bibinfo {volume} {151}},\ \bibinfo {pages} {124103} (\bibinfo
  {year} {2019})}\BibitemShut {NoStop}%
\bibitem [{\citenamefont {Tuckerman}(2010)}]{TuckermanBook}%
  \BibitemOpen
  \bibfield  {author} {\bibinfo {author} {\bibfnamefont {M.~E.}\ \bibnamefont
  {Tuckerman}},\ }\href@noop {} {\emph {\bibinfo {title} {Statistical
  Mechanics: Theory and Molecular Simulation}}}\ (\bibinfo  {publisher} {Oxford
  University Press},\ \bibinfo {year} {2010})\BibitemShut {NoStop}%
\bibitem [{\citenamefont {Frenkel}\ and\ \citenamefont
  {Smit}(2002)}]{DaanFrenkel_book}%
  \BibitemOpen
  \bibfield  {author} {\bibinfo {author} {\bibfnamefont {D.}~\bibnamefont
  {Frenkel}}\ and\ \bibinfo {author} {\bibfnamefont {B.}~\bibnamefont {Smit}},\
  }\href@noop {} {\emph {\bibinfo {title} {Understanding Molecular Simulation:
  From Algorithms to Applications}}}\ (\bibinfo  {publisher} {Academic Press,
  San Diego, USA},\ \bibinfo {year} {2002})\BibitemShut {NoStop}%
\bibitem [{\citenamefont {Zarotiadis}(2018)}]{RhiannonThesis}%
  \BibitemOpen
  \bibfield  {author} {\bibinfo {author} {\bibfnamefont {R.}~\bibnamefont
  {Zarotiadis}},\ }\href@noop {} {Master's thesis},\ \bibinfo  {school} {ETH
  Zurich} (\bibinfo {year} {2018})\BibitemShut {NoStop}%
\bibitem [{\citenamefont {Meyer}(1970)}]{Meyer1970DVR}%
  \BibitemOpen
  \bibfield  {author} {\bibinfo {author} {\bibfnamefont {R.}~\bibnamefont
  {Meyer}},\ }\href@noop {} {\bibfield  {journal} {\bibinfo  {journal}
  {J.~Chem. Phys.}\ }\textbf {\bibinfo {volume} {52}},\ \bibinfo {pages} {2053}
  (\bibinfo {year} {1970})}\BibitemShut {NoStop}%
\bibitem [{\citenamefont {Light}, \citenamefont {Hamilton},\ and\ \citenamefont
  {Lill}(1985)}]{Light1985DVR}%
  \BibitemOpen
  \bibfield  {author} {\bibinfo {author} {\bibfnamefont {J.~C.}\ \bibnamefont
  {Light}}, \bibinfo {author} {\bibfnamefont {I.~P.}\ \bibnamefont {Hamilton}},
  \ and\ \bibinfo {author} {\bibfnamefont {J.~V.}\ \bibnamefont {Lill}},\
  }\href {\doibase 10.1063/1.448462} {\bibfield  {journal} {\bibinfo  {journal}
  {J.~Chem. Phys.}\ }\textbf {\bibinfo {volume} {82}},\ \bibinfo {pages} {1400}
  (\bibinfo {year} {1985})}\BibitemShut {NoStop}%
\bibitem [{\citenamefont {Bunkin}\ and\ \citenamefont
  {Tugov}(1973)}]{Bunkin1973Morse}%
  \BibitemOpen
  \bibfield  {author} {\bibinfo {author} {\bibfnamefont {F.}~\bibnamefont
  {Bunkin}}\ and\ \bibinfo {author} {\bibfnamefont {I.}~\bibnamefont {Tugov}},\
  }\href {\doibase 10.1103/PhysRevA.8.601} {\bibfield  {journal} {\bibinfo
  {journal} {Phys. Rev. A}\ }\textbf {\bibinfo {volume} {8}},\ \bibinfo {pages}
  {601} (\bibinfo {year} {1973})}\BibitemShut {NoStop}%
\bibitem [{Note6()}]{Note6}%
  \BibitemOpen
  \bibinfo {note} {Supplementary material at xxxxx}\BibitemShut {NoStop}%
\bibitem [{\citenamefont {Andersen}(1983)}]{RATTLE}%
  \BibitemOpen
  \bibfield  {author} {\bibinfo {author} {\bibfnamefont {H.~C.}\ \bibnamefont
  {Andersen}},\ }\href {\doibase 10.1016/0021-9991(83)90014-1} {\bibfield
  {journal} {\bibinfo  {journal} {J.~Comput. Phys.}\ }\textbf {\bibinfo
  {volume} {52}},\ \bibinfo {pages} {24} (\bibinfo {year} {1983})}\BibitemShut
  {NoStop}%
\bibitem [{Note7()}]{Note7}%
  \BibitemOpen
  \bibinfo {note} {In fact, we show in the SI that the unconstrained ensemble
  remains far from either instanton regardless of the $\lambda $ value
  used.}\BibitemShut {Stop}%
\bibitem [{Note8()}]{Note8}%
  \BibitemOpen
  \bibinfo {note} {Note that the instanton orbits of golden-rule rate theory
  [Ref.~\protect \rev@citealpnum {GoldenRPI}] do not have a zero-frequency
  permutational mode like the instantons for adiabatic reactions [Ref.~\protect
  \rev@citealpnum {RPInst}]}\BibitemShut {NoStop}%
\bibitem [{\citenamefont {Rips}\ and\ \citenamefont
  {Pollak}(1995)}]{Rips1995ET}%
  \BibitemOpen
  \bibfield  {author} {\bibinfo {author} {\bibfnamefont {I.}~\bibnamefont
  {Rips}}\ and\ \bibinfo {author} {\bibfnamefont {E.}~\bibnamefont {Pollak}},\
  }\href {\doibase 10.1063/1.470209} {\bibfield  {journal} {\bibinfo  {journal}
  {J.~Chem. Phys.}\ }\textbf {\bibinfo {volume} {103}},\ \bibinfo {pages}
  {7912} (\bibinfo {year} {1995})}\BibitemShut {NoStop}%
\bibitem [{\citenamefont {Landau}(1932)}]{Landau1932LZ}%
  \BibitemOpen
  \bibfield  {author} {\bibinfo {author} {\bibfnamefont {L.~D.}\ \bibnamefont
  {Landau}},\ }\href@noop {} {\bibfield  {journal} {\bibinfo  {journal} {Phys.
  Z. Sowjetunion}\ }\textbf {\bibinfo {volume} {2}},\ \bibinfo {pages} {46}
  (\bibinfo {year} {1932})}\BibitemShut {NoStop}%
\bibitem [{\citenamefont {Zener}(1932)}]{Zener1932LZ}%
  \BibitemOpen
  \bibfield  {author} {\bibinfo {author} {\bibfnamefont {C.}~\bibnamefont
  {Zener}},\ }\href {\doibase 10.1098/rspa.1932.0165} {\bibfield  {journal}
  {\bibinfo  {journal} {Proc. R. Soc. Lond. A}\ }\textbf {\bibinfo {volume}
  {137}},\ \bibinfo {pages} {696} (\bibinfo {year} {1932})}\BibitemShut
  {NoStop}%
\bibitem [{\citenamefont {Kapil}, \citenamefont {Behler},\ and\ \citenamefont
  {Ceriotti}(2016)}]{Kapil2016PI}%
  \BibitemOpen
  \bibfield  {author} {\bibinfo {author} {\bibfnamefont {V.}~\bibnamefont
  {Kapil}}, \bibinfo {author} {\bibfnamefont {J.}~\bibnamefont {Behler}}, \
  and\ \bibinfo {author} {\bibfnamefont {M.}~\bibnamefont {Ceriotti}},\ }\href
  {\doibase 10.1063/1.4971438} {\bibfield  {journal} {\bibinfo  {journal}
  {J.~Chem. Phys.}\ }\textbf {\bibinfo {volume} {145}},\ \bibinfo {pages}
  {234103} (\bibinfo {year} {2016})}\BibitemShut {NoStop}%
\bibitem [{\citenamefont {Earl}\ and\ \citenamefont {Deem}(2005)}]{B509983H}%
  \BibitemOpen
  \bibfield  {author} {\bibinfo {author} {\bibfnamefont {D.~J.}\ \bibnamefont
  {Earl}}\ and\ \bibinfo {author} {\bibfnamefont {M.~W.}\ \bibnamefont
  {Deem}},\ }\href {\doibase 10.1039/B509983H} {\bibfield  {journal} {\bibinfo
  {journal} {Phys. Chem. Chem. Phys.}\ }\textbf {\bibinfo {volume} {7}},\
  \bibinfo {pages} {3910} (\bibinfo {year} {2005})}\BibitemShut {NoStop}%
\bibitem [{\citenamefont {Peng}\ \emph {et~al.}(2014)\citenamefont {Peng},
  \citenamefont {Cao}, \citenamefont {Zhou},\ and\ \citenamefont
  {Voth}}]{doi:10.1021/ct500447r}%
  \BibitemOpen
  \bibfield  {author} {\bibinfo {author} {\bibfnamefont {Y.}~\bibnamefont
  {Peng}}, \bibinfo {author} {\bibfnamefont {Z.}~\bibnamefont {Cao}}, \bibinfo
  {author} {\bibfnamefont {R.}~\bibnamefont {Zhou}}, \ and\ \bibinfo {author}
  {\bibfnamefont {G.~A.}\ \bibnamefont {Voth}},\ }\href {\doibase
  10.1021/ct500447r} {\bibfield  {journal} {\bibinfo  {journal} {J. Chem.
  Theory Comput.}\ }\textbf {\bibinfo {volume} {10}},\ \bibinfo {pages} {3634}
  (\bibinfo {year} {2014})}\BibitemShut {NoStop}%
\end{thebibliography}%

\clearpage
~~~~~~~~~~~~~~~~~~~~~~~~~~~~~~~\\


\section*{Supplementary material}

# Supporting information: Nonadiabatic quantum transition-state theory in the golden-rule limit. II. Overcoming the pitfalls of the saddle-point and semiclassical approximations

Wei Fang,[1] Manish J. Thapa,[1] and Jeremy O. Richardson[1, *]

[1]*Laboratory of Physical Chemistry, ETH Zurich, 8093 Zurich, Switzerland*

(Dated: October 9, 2019)

## Abstract

In this supporting information we provide further details of the simulations reported in the manuscript and additional discussions in support of the conclusions reached.

* jeremy.richardson@phys.chem.ethz.ch

## S.I. WOLYNES FREE-ENERGY INTEGRATION

In this section, we report the free-energy derivative data we obtained in the Wolynes (unconstrained) simulations for both system I and II (Figs. S1–S3). One can see that all of the free-energy derivative curves make a sharp turn near $\lambda^*$ (where $\frac{dF_u}{d\lambda}\big|_{\lambda^*} = 0$). This occurs when the $\phi_u(\tau)$ curve changes suddenly from being dominated by one TS to the other and is a sign that Wolynes theory is likely to break down.

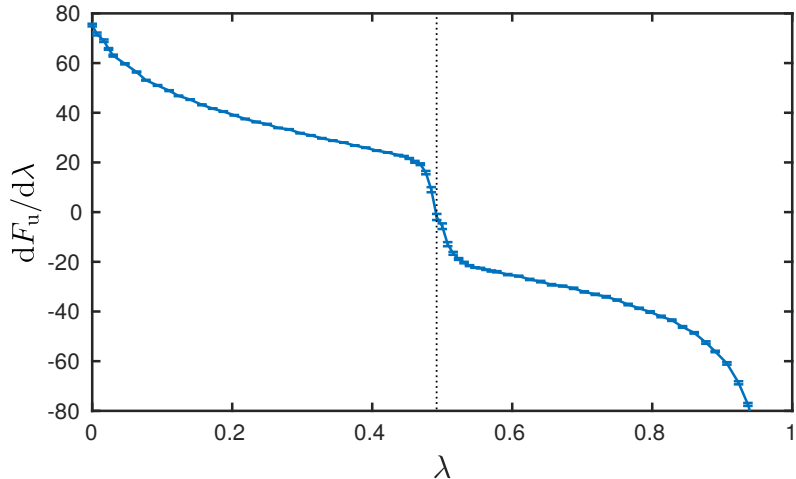

FIG. S1. Unconstrained free-energy integrand for system I at $\beta = 3$. The dashed line shows the value of $\lambda$ at which the free-energy derivative passes through zero, which defines the optimal $\lambda^*$ used in both Wolynes theory and the GR-QTST method under our new ansatz.

Another way of thinking about this problem of Wolynes theory is that the thermodynamical average of these samples gives $\frac{dF_u}{d\lambda} = 0$. By decomposing the contributions to $\frac{dF_u}{d\lambda}$ (Fig. S4), one notices that samples from the bottom crossing contribute to a positive $\frac{dF_u}{d\lambda}$, while the top crossing contributes to a negative $\frac{dF_u}{d\lambda}$. This means that because the two TS have very different optimal values of $\lambda$, Wolynes theory ends up picking a bad $\lambda^*$ value that tries to compromise between the two TS.

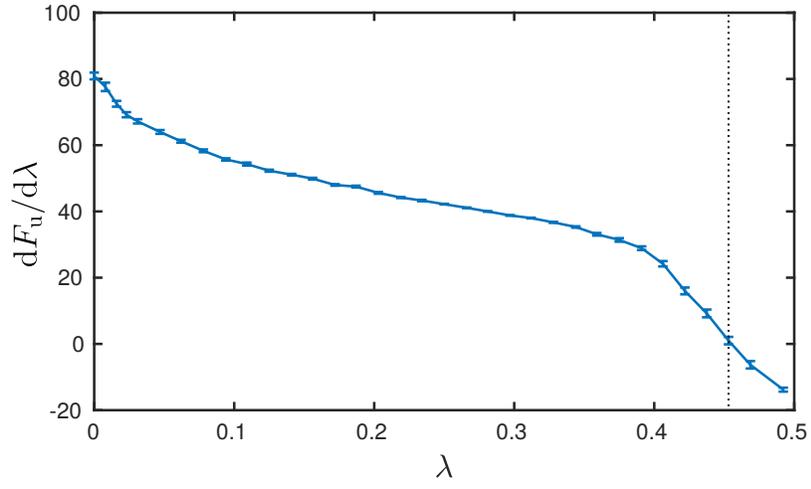

FIG. S2. As Fig. S1, for system I with $\beta = 1$.

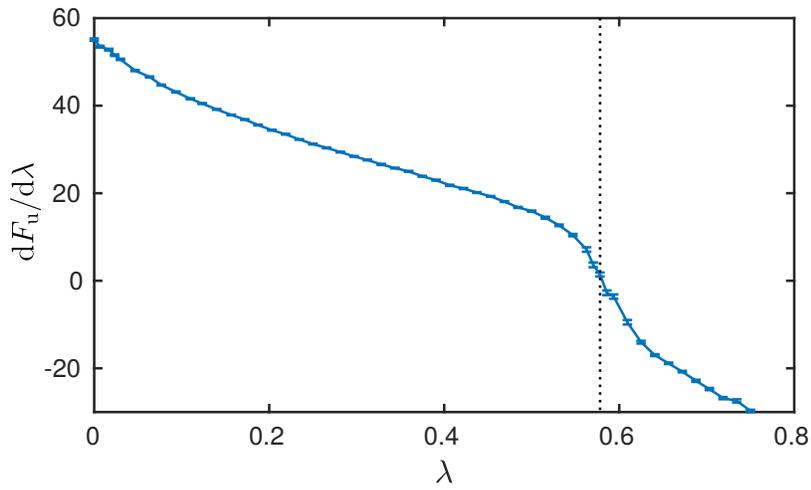

FIG. S3. As Fig. S1, for system II with $\beta = 3$.

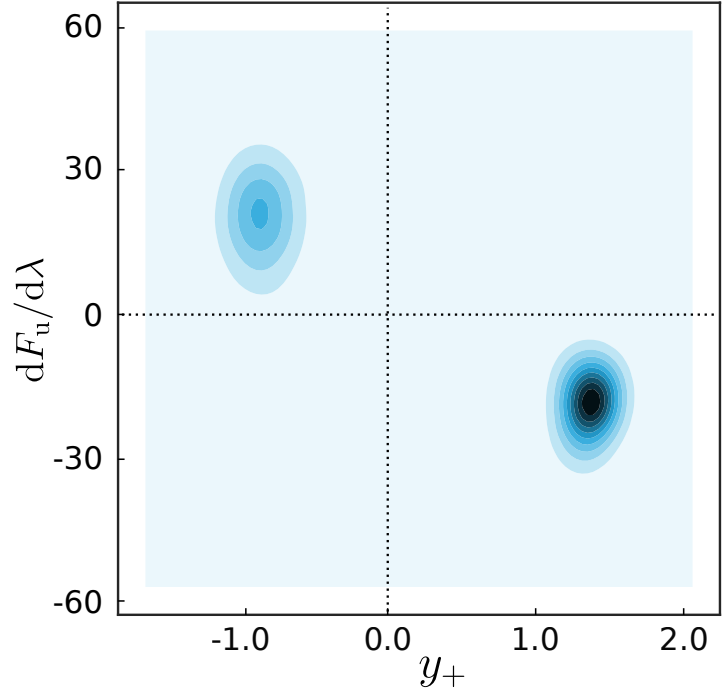

FIG. S4. Density plot of the distributions of $\frac{dF_u}{d\lambda}$ and $y_+$ in the unconstrained ensemble used for Wolynes theory at its optimal value of $\lambda = \lambda^* = 0.49$, for system I at $\beta = 3$.

## S.II. GR-QTST FREE-ENERGY INTEGRATION

In this section, we report the free-energy integrand data we obtained in the GR-QTST simulations for both system I and II (Figs. S5 and S6). One can see that the integrands are well-behaved after the coordinate transformation to $\xi$.

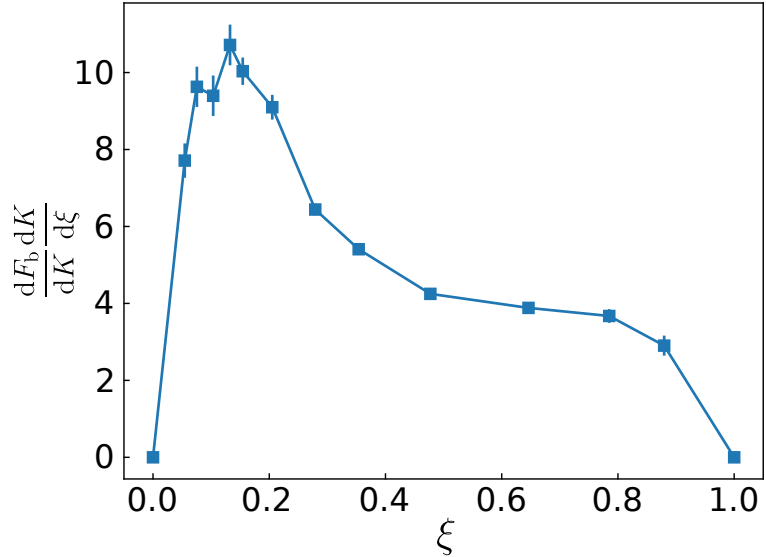

FIG. S5. The GR-QTST integrand at $\lambda = 0.45$, the optimal $\lambda^*$ value used in Wolynes theory, for system I at $\beta = 1$.

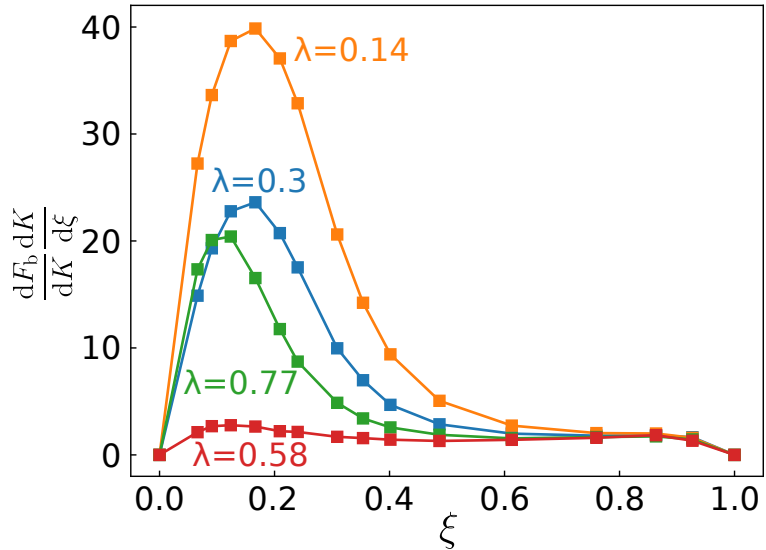

FIG. S6. The GR-QTST integrand at four different $\lambda$ values for system II at $\beta = 3$. The statistical error bars are smaller than the symbol size.

## S.III. UNCONSTRAINED VS CONSTRAINED SAMPLING

In this section, we show a comparison of the regions sampled in unconstrained simulations and constrained simulations (Fig. S7). Instanton-like configurations, i.e. those with the correct $x_+$ and $y_+$ values, are sampled with the GR-QTST energy constraint regardless of the value of $\lambda$, whereas unconstrained sampling misses them completely.

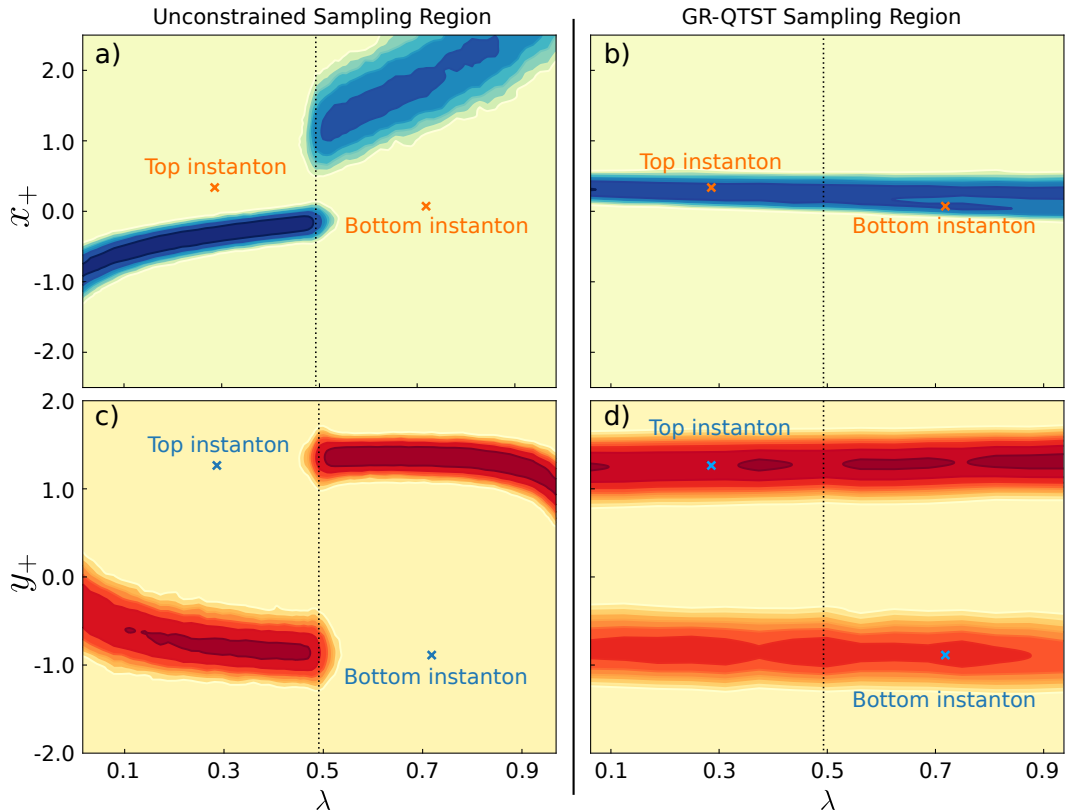

FIG. S7. Density plot of a relative free energy, defined as $F_\lambda(x_+) = -\log p_\lambda(x_+)$, over a range of $\lambda$ values, for system I at $\beta = 3$ sampled in (a) unconstrained simulations, and (b) sampled with the GR-QTST energy constraint. $p_\lambda(x_+)$ is the probability density of $x_+$ sampled by unconstrained or constrained path-integral molecular dynamics simulations at a given $\lambda$. The density is normalized at each value of $\lambda$ such that $\int p_\lambda(x_+) \mathrm{d}x_+ = 1$. (c) and (d) are the same as (a) and (b) except that the densities of $y_+$ are measured instead. Green in (a) and (b) or yellow in (c) and (d) marks the regions that are rarely sampled. The dotted lines indicate $\lambda^*$. The instantons are also marked on the phase space.


% \begin{figure}
% \centering
% \includegraphics[page=-]{SI.pdf}
% \end{figure}

\end{document}